\documentclass[a4paper,12pt]{article}

\usepackage [T1]{fontenc}
\usepackage {calligra}
\usepackage [T1]{pbsi}
\usepackage{CJKutf8} 
\usepackage[colorlinks=true,linkcolor=blue]{hyperref}
\usepackage{amsmath,amssymb,amsfonts} 
\usepackage{graphicx}                 
\usepackage{color}                    
\usepackage{hyperref}                 
\usepackage{cite}

\definecolor{red}{rgb}{1,0,0}           
\definecolor{green}{rgb}{0,1,0}
\definecolor{blue}{rgb}{0,0,1}
\definecolor{darkblue}{rgb}{0,0,0.5}
\definecolor{lightblue}{rgb}{.5,.5,1}
\definecolor{lightgray}{gray}{.87}          
\definecolor{Dark}{gray}{.20}
\definecolor{pink}{rgb}{.95,0.82,0.92}  
\definecolor{yellow}{rgb}{1,1,0}
\definecolor{lightyellow}{rgb}{1,1,.5}
\definecolor{purple}{rgb}{0.7,0,0.85}
\definecolor{darkgreen}{rgb}{0,0.5,0}
\definecolor{orange}{rgb}{0.8,0.2,0.2}
\def \be {\begin{equation}}
\def \ee {\end{equation}}
\def \bea {\begin{eqnarray}}
\def \eea {\end{eqnarray}}
\def \nn {\nonumber}

\def \rr {\raise.35ex\hbox{\small $\prime$}\kern-.17em{\mbox{\large $\imath$}}}
\def \del {\partial}
\def \dels {\partial\kern-.5em / \kern.5em}
\def \As {{A\kern-.5em / \kern.5em}}
\def \Ds {D\kern-.7em / \kern.5em}

\def \a {\alpha}

\def \eps {\epsilon}

\def \th {\theta}

\newcommand{\detail}[1]{}

\newcommand{\Misconception}[2]{
\vskip1em
\noindent {\bf Misconception #1}
\begin{quotation}
{\em
#2
}
\end{quotation}
}

\newcommand{\hide}[1]{}

\begin{document}

\pagestyle{plain}

\begin{CJK}{UTF8}{bsmi}

\begin{titlepage}

\begin{center}

\noindent
\textbf{\LARGE
Asymptotic Black Holes \\
}
\vskip .5in
{\large 
Pei-Ming Ho
\footnote{e-mail address: pmho@phys.ntu.edu.tw},
}
\\
{\vskip 10mm \sl

Department of Physics and Center for Theoretical Sciences, \\
Center for Advanced Study in Theoretical Sciences, \\
National Taiwan University, Taipei 106, Taiwan,
R.O.C. 
}\\

\vskip 3mm
\vspace{60pt}
\begin{abstract}

Following earlier works on the KMY model of 
black-hole formation and evaporation,
we construct the metric for a matter sphere in gravitational collapse,
with the back-reaction of pre-Hawking radiation taken into consideration.
The mass distribution and collapsing velocity of the matter sphere
are allowed to have an arbitrary radial dependence.
We find that a generic gravitational collapse asymptote to 
a universal configuration which resembles a black hole
but without horizon.
This approach clarifies several misunderstandings about
black-hole formation and evaporation,
and provides a new model
for black-hole-like objects in the universe.

\end{abstract}
\end{center}
\end{titlepage}

\setcounter{page}{1}
\setcounter{footnote}{0}
\setcounter{section}{0}


\section{Introduction}

In the study of the formation and evaporation of black holes,
various approximation schemes have often been adopted.
In particular,
as the surface of a collapsing star gets very close to the Schwarzschild radius,
pre-Hawking radiation is expected to arise before the horizon emerges,
but its back-reaction is conventionally ignored.
We will show in this paper
the significance of this back-reaction on geometry
despite its extreme weakness.

A new approach to black-hole formation and evaporation
was proposed by Kawai, Matsuo and Yokokura in 2013 \cite{Kawai:2013mda}.
It allows us to study the space-time geometry for a collapsing sphere,
{\em without} ignoring the back-reaction of pre-Hawking radiation
\footnote{
In previous works 
\cite{Kawai:2013mda,Kawai:2014afa,Ho:2015fja,Kawai:2015uya,Ho:2015vga},
``pre-Hawking radiation'' was simply called ``Hawking radiation''.
}
as a solution to the Einstein equation.
It was found that
the back-reaction of pre-Hawking radiation
prevents the apparent horizon to emerge
whenever there is complete evaporation
\cite{Kawai:2013mda,Kawai:2014afa,Ho:2015fja,Kawai:2015uya,Ho:2015vga}.
\footnote{
We are interested in the gravitational collapse of astronomical objects.
If the formula for pre-Hawking radiation needs to be modified at short distance
such that the evaporation leaves a remnant,
basic ideas about the KMY model persist
and necessary modifications are straightforward \cite{Ho:2015vga}.
}
The proposal that pre-Hawking radiation prevents
the formation of horizon was also proposed in Refs.
\cite{Vachaspati:2006ki,Saini:2015dea,Mersini-Houghton:2014zka,Mersini-Houghton:2014cta}.

Various aspects of this new approach,
including its generalization
and its connection to the information loss paradox,
have been discussed in the literature
\cite{Kawai:2013mda,Kawai:2014afa,Ho:2015fja,Kawai:2015uya,Ho:2015vga}.
While it was the special case of a matter sphere 
collapsing at the speed of light that was
considered in the original paper \cite{Kawai:2013mda},
we shall refer to all models of black-hole formation and evaporation
that are constructed following this new approach
\cite{Kawai:2013mda,Kawai:2014afa,Ho:2015fja,Kawai:2015uya,Ho:2015vga}
as the KMY model.

In this paper,
we focus on the geometry inside the collapsing sphere,
generalizing the results of earlier papers
\cite{Kawai:2013mda,Kawai:2014afa,Kawai:2015uya}.
The metric of the full space-time is found
for mass densities and collapsing velocities 
with arbitrary radial dependence.
It allows us to clarify various misunderstandings 
about black-hole formation and evaporation.
Furthermore, 
we shall emphasize the notion of the asymptotic black hole,
which refers to a universal asymptotic state of late-stage gravitational collapse.
A generic gravitational collapse
with a sufficiently large mass
eventually evolves into a configuration
which is almost indistinguishable from this asymptotic state,
and, 
for a distant observer,
it is also very difficult to distinguish it from a real black hole with horizon.

In Sec.\ref{KMY-Model},
we first review the KMY model,
and then derive the red-shift factor inside a collapsing sphere
for generic radial distributions of mass and velocity.
The space-time metric for the collapsing sphere is given in Sec.\ref{Geometry}.
We verify that the energy-momentum tensor 
involves no singularity at the surface of the collapsing sphere,
and analyze the case of a sphere collapsing at the speed of light
in more detail as an example.
Finally,
we clarify misconceptions in Sec.\ref{Misconceptions},
and comment on implications of the KMY model in Sec.\ref{Comments}.

\section{KMY Model}
\label{KMY-Model}

The KMY model \cite{Kawai:2013mda} describes
a gravitational collapse
with the back-reaction of pre-Hawking radiation taken into consideration.
The major difference between the KMY model 
and the conventional model of black holes is the following.
In the conventional model,
it is assumed that Hawking radiation
(as well as pre-Hawking radiation)
exists only at distance,
and the neighborhood of the horizon is a vacuum state.
However, 
it was shown \cite{Almheiri:2012rt,Braunstein} 
that this is not a consistent assumption.
Even if the horizon is in vacuum in the beginning,
it will evolve into the state of a ``firewall'' at a later time.
On the other hand,
the KMY model assumes that 
the pre-Hawking radiation contributes to the energy-momentum tensor
in the Einstein equation
even at the surface of the collapsing sphere,
when it is close to the Schwarzschild radius.
Taking into account of the back-reaction of the pre-Hawking radiation,
one can show
\cite{Kawai:2013mda,Kawai:2014afa,Ho:2015fja,Kawai:2015uya,Ho:2015vga}
that the whole collapsing sphere can evaporate completely 
without apparent horizon.

In this paper,
we shall make the following assumptions for simplicity
and for technical convenience.
We shall assume
(1) spherical symmetry and
(2) pre-Hawking radiation being composed of nothing but massless dust
(spin-0 free particles).

\subsection{Geometry Outside Collapsing Sphere}

In this subsection,
we review the geometry outside a collapsing sphere.
Please refer to 
Refs.\cite{Kawai:2013mda,Ho:2015fja,Ho:2015vga} 
for more details.

For the space-time geometry with spherical symmetry,
one can always define the radial coordinate $r$ by demanding that
the metric is of the form
\be
ds^2 = g_{tt}(t, r) dt^2 + 2 g_{tr}(t, r) dt dr + g_{rr}(t, r) dr^2 + r^2 d\Omega^2,
\ee
where $d\Omega^2 = d\th^2 + \sin^2\th d\phi^2$,
so that the area of a sphere of radius $r$ is $4\pi r^2$.

Denote by $R_0(u)$ the radial coordinate of 
the surface of the collapsing sphere.
\footnote{
In the case when there is an everlasting collapse without boundary,
we have $R_0(u) = \infty$,
and most of the results of this paper still apply.
}
There is nothing but pre-Hawking radiation 
(which is assumed to be composed of massless spin-0 particles)
for $r > R_0(u)$.
It follows from the Einstein equation 
\be
G_{\mu\nu} = 8\pi G T_{\mu\nu}
\ee
that the geometry for $r > R_0(u)$ should be described by the outgoing Vaidya metric 
\footnote{
See Appendix A for more about the outgoing and ingoing Vaidya metrics.
}
\cite{Vaidya:1951zz}:
\be
ds^2 = - \left( 1 - \frac{a_0(u)}{r} \right) du^2
- 2 du dr + r^2 d\Omega^2
\qquad (r \geq R_0(u)),
\label{OGVM}
\ee
where $a_0(u)$ is the Schwarzschild radius.

According to the Einstein equation,
the energy-momentum tensor for pre-Hawking radiation
outside the collapsing sphere is
\be
T_{uu} = - \frac{1}{8\pi G} \frac{\dot{a}_0(u)}{r^2}
\qquad (r \geq R_0(u)),
\label{EMT-outgoing-Vaidya-0}
\ee
with all other components of the energy-momentum tensor
($T_{ur}, T_{rr}, T_{\th\th}, T_{\phi\phi}$) vanishing.
As $T_{uu}$ should be positive to represent an outgoing radiation,
$\dot{a}_0(u)$ should be negative.

The Schwarzschild radius $a_0(u)$ would correspond to 
the location of a (white-hole) apparent horizon
only if $R_0(u)$ becomes smaller than $a_0(u)$,
since the outgoing Vaidya metric only applies to $r > R_0(u)$.
This is however impossible if there is complete evaporation
\cite{Kawai:2013mda,Ho:2015fja,Ho:2015vga}.
The argument goes as follows.
\footnote{
For more detailed and comprehensive explanation,
see Refs.\cite{Kawai:2013mda,Ho:2015fja,Ho:2015vga}.
}
If there would be an event horizon,
there must be some infalling light-like trajectories 
which are geodesically incomplete from the viewpoint of a distant observer
(in the coordinate patch of $(u, r)$).
If there would be an apparent (but not event) horizon,
there must be a moment when certain infalling light-like trajectories cross over $a_0(u)$.
There is no event nor apparent horizon 
if all infalling light-like geodesics do not cross over $a_0(u)$
and they are all geodesically complete.

It can be shown that $a_0(u)$ shrinks at a speed faster than light
as long as $\dot{a}_0 < 0$,
it is thus impossible for infalling light-like geodesics to cross over $a_0(u)$
within the coordinate patch of $(u, r)$.
Furthermore,
if there is complete evaporation
\be
a_0(u) = 0 
\qquad \forall u \geq u^*
\label{CE}
\ee
for some finite $u^*$,
the outgoing Vaidya space-time turns into the Minkowski space-time at $u = u^*$.
As the Minkowski space-time is geodesically complete,
all infalling trajectories can be shown to be complete
\cite{Kawai:2013mda,Ho:2015fja,Ho:2015vga}.
The claim that there is no horizon in gravitational collapse
even for an astronomically massive star, 
due to the creation of quantum particles,
can be unsettling at first.
But it is supported by explicit calculation,
and we will explain more below in Sec.\ref{Misconceptions} and the appendix.

Let us also emphasize here that
the assumption of complete evaporation is not necessary 
for most of the discussion below.
If there would be a black-hole event horizon,
it just means that the coordinate system $(u, r, \th, \phi)$
(and our discussion based on this coordinate system)
can only apply to the space-time outside the event horizon,

The trajectories of $R_0(u)$ and $a_0(u)$ depend on
further details about the collapsing sphere.
In the extreme case of gravitational collapse
where the matter sphere collapses at the speed of light,
the trajectory of $R_0(u)$ should be an ingoing light-like geodesic
\cite{Kawai:2013mda}.
Due to the continuity of the metric at $r = R_0(u)$,
the outgoing Vaidya metric (\ref{OGVM}) implies that
\be
\frac{dR_0(u)}{du} = - \frac{1}{2} \left( 1 - \frac{a_0(u)}{R_0(u)} \right).
\label{R-trajectory}
\ee

When the separation between $R_0(u)$ and $a_0(u)$
is much shorter than the wavelength of the dominant mode
in pre-Hawking radiation,
the energy flux in pre-Hawking radiation 
should be well approximated by the usual formula for Hawking radiation,
\be
\frac{da_0(u)}{du} \simeq - \frac{\sigma}{a_0^2(u)}
\label{a-trajectory}
\ee
for large $a_0(u)$
(See Ref.\cite{Kawai:2013mda}
and Sec. \ref{HawkingRadiation}.),
which leads to a slow change in the Bondi mass $M_0(u) = a_0(u)/2$.

The constant $\sigma$ is given by
\be
\sigma \simeq \frac{N_0\ell_p^2}{48\pi},
\label{sigma}
\ee
where $N_0$ is the number of species of massless particles.
This equation implies that 
\be
a_0(u) = \left\{
\begin{array}{ll}
(3\sigma(u^* - u))^{1/3} & (u < u^*),
\\
0 & (u > u^*).
\end{array}
\right.
\label{a-u-1/3}
\ee
There is thus complete evaporation,
up to the possibility of a small remnant 
as eq.(\ref{a-trajectory}) is valid only for large $a_0$.
We shall focus on large-scale physics in this paper,
and view the remnant as one of the particles in pre-Hawking radiation.

Given the solution (\ref{a-u-1/3}) of $a_0(u)$,
eq.(\ref{R-trajectory}) then determines,
for the collapse at the speed of light,
the trajectory of $R_0(u)$,
which sets the boundary of the outgoing Vaidya metric (\ref{OGVM}).

\subsection{Red-Shift Factor}
\label{Collapse}

\begin{figure}
\vskip-2em
\begin{minipage}[c]{0.67\textwidth}
\vskip-4em
\center
\includegraphics[scale=0.4,bb=0 100 500 500]{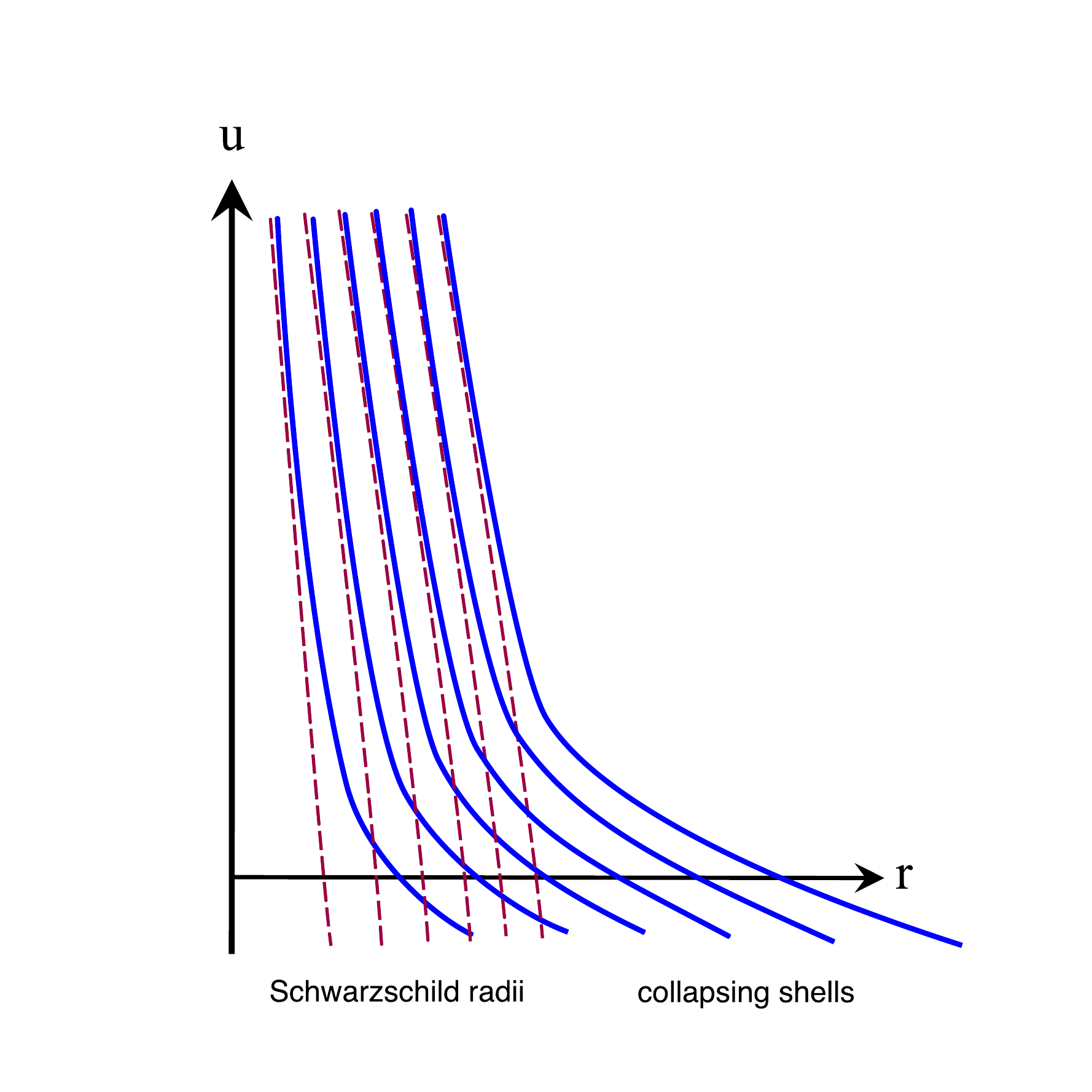}
\vskip2em
\end{minipage}
\begin{minipage}[c]{0.35\textwidth}
\caption{\small
Thin matter shells (blue curves) in gravitational collapse approach to, 
and then stay close to the Schwarzschild radii (red dash curves) of each thin shell.
The space between neighboring matter shells contains pre-Hawking radiation
and is described by the outgoing Vaidya metric.
}
\label{CollapsingShells}
\end{minipage}
\vskip1em
\end{figure}

Conventionally,
it is believed that (pre-)Hawking radiation 
can be ignored when one considers
the geometry inside a collapsing star.
A more reliable approach was
proposed in Refs.\cite{Kawai:2013mda,Kawai:2014afa},
without neglecting the back-reaction of (pre-)Hawking radiation.
In this approach,
the interior of the collapsing sphere is
decomposed into infinitely many infinitesimally thin matter shells
separated by infinitesimally thin layers of space 
in which there is nothing but pre-Hawking radiation.
(See Fig.\ref{CollapsingShells}.)
Due to spherical symmetry,
one can use the outgoing Vaidya metric for each layer of space,
and then take the continuum limit to find the continuous metric.
The crucial step is to compute the red-shift factor relating 
the Eddington retarded times in different layers.
This is what we will do in this subsection.

Let us first discretize the collapsing sphere into $N$ thin shells.
Consider the $n$-th collapsing thin shell of radius $R_n(u_n)$.
(The outermost shell is labelled by $n = 0$.)
The Eddington retarded time coordinates $u_{n}$ and $u_{n+1}$
in the two thin layers of space sandwiching this collapsing matter shell 
are related by a red-shift factor that can be determined 
by the continuity of the induced metric at $r = R_n(u_n)$:
\be
\left(1 - \frac{a_n(u_n)}{R_n(u_n)}\right) du_n^2 + 2 du_n dR_n(u_n)
= \left(1 - \frac{a_{n+1}(u_{n+1})}{R_n(u_n)}\right) du_{n+1}^2 + 2 du_{n+1} dR_n(u_n).
\ee
For a given trajectory $R_n(u_n)$,
we find the red-shift factor
\bea
\frac{du_{n+1}}{du_n} 
&=& 
\frac{1}{\left(1-\frac{a_{n+1}}{R_n}\right)}
\left[
\sqrt{
\left(\frac{dR_n}{du_n}\right)^2
+ \left(1-\frac{a_{n+1}}{R_n}\right)
\left[
\left(1 - \frac{a_n}{R_n}\right) + 2 \frac{dR_n}{du_n}
\right]
}
- \frac{dR_n}{du_n}
\right]
\nn\\
&=&
1+\frac{1}{R_n(u_n)-a_n(u_n)+R_n(u_n)\dot{R}_n(u_n)}\frac{da_n(u_n)}{2}
+ {\cal O}(da_n^2),
\eea
where $\dot{R}_n \equiv \frac{dR_n}{du_n}$.
Here $da_n(u_n) \equiv a_{n+1}(u_{n+1}) - a_n(u_n)$
is assumed to be very small so that higher orders of $da_n$ can be ignored.
It follows that the red-shift factor between 
any two (not necessarily neighboring) layers of space with labels $m$ and $n$ ($m > n$) is
\be
\frac{du_m}{du_n} = \prod_{k=n}^{m-1} \left( 1+\frac{1}{R_k-a_k+R_k\dot{R}_k}\frac{da_k}{2} \right).
\ee

In the limit of infinite $N$,
we use a continuous parameter $\a$ to replace the index $n$,
and the expression above becomes
\be
\frac{du_{\a_2}}{du_{\a_1}} = 
e^{\frac{1}{2} \int_{\a_1}^{\a_2} d\a \; \frac{da_{\a}(u_{\a})}{d\a} \frac{1}{R_{\a}(u_{\a}) - a_{\a}(u_{\a})+R_{\a}(u_{\a})\dot{R}_{\a}(u_{\a})}}.
\label{red-shift-0}
\ee
The gives the red-shift factor between any two layers of space
in the continuum limit.
This formula is valid for arbitrary trajectories of the thin shells,
assuming that the ordering of the shells is preserved.
It allows us to think of the coordinate $u_{\a_1}$ in a layer
as a function of the coordinate $u_{\a_2}$ in another layer.
We shall reserve the symbol $u$ with no argument 
as the Eddington retarded time for $r > R_0(u)$ 
(i.e. $u = u_0$),
and the coordinate $u_{\a}(u)$ on any shell can be viewed as a function of $u$
through the red-shift factor.

Taking a step further and using the radial coordinate $r$ to replace the parameter $\alpha$,
we introduce two functions $a(u, r)$ and $V(u, r)$,
where $a(u, r)$ is defined by $a(u, r) = a_{\a}(u_{\a}(u))$
and $V(u, r)$ is defined by $V(u, r) = \frac{dR_{\a}(u_{\a})}{du_{\a}}$
when $r = R_{\a}(u_{\a}(u))$ for a certain value of $\alpha$.
It is assumed that,
at a given time $u$,
there is a unique thin shell labelled by $\a$ that is of radius $r$.

In terms of $a(u, r)$ and $V(u, r)$,
the red-shift factor between the Eddington retarded time $u(r)$ at a given coordinate $r$
(That is, $u(r) = u_{\a}$ if $r = R_{\a}$.)
and the Eddington retarded time $u = u(R_0)$ outside the collapsing sphere is 
\be
e^{\psi(u, r)} \equiv \frac{du(r)}{du} = e^{-\frac{1}{2}\int_r^{R_0(u)} dr' \; \frac{a'(u, r')}{r' - a(u, r') + r' V(u, r')}},
\label{red-shift-general}
\ee
where 
$a' \equiv \frac{\del a}{\del r}$.

In the case when all thin shells are collapsing at the speed of light,
\be
\dot{R}_{\a}(u_{\a}) \equiv \frac{dR_{\a}(u_{\a})}{du_{\a}} = - \frac{1}{2}\left( 1 - \frac{a_{\a}(u_{\a})}{R_{\a}(u_{\a})} \right),
\label{R-alpha}
\ee
hence
\be
V(u, r) = - \frac{1}{2}\left( 1 - \frac{a(u, r)}{r} \right),
\label{V-lightlike}
\ee
the expression for the red-shift factor can be simplified as
\be
e^{\psi(u, r)}
= e^{-\int_{r}^{R_0(u)} dr' \; \frac{a'(u, r')}{r' - a(u, r')}}.
\label{red-shift}
\ee
This was first obtained in Ref.\cite{Kawai:2013mda}.
Eq.(\ref{red-shift-general}) is a generalization of this expression,
and is the main result of this subsection.

\subsection{Trajectories of Collapsing Shells}
\label{Traj}

The geometry inside the collapsing sphere depends on
the trajectories of each thin shell $R_{\a}$.
If all thin shells are collapsing at the speed of light,
the trajectories of $R_{\a}$ are given by eq.(\ref{R-alpha}).
More generally, 
it could be determined by an equation of motion of the form
\be
\frac{d^2{R}_{\a}}{du^2_{\a}} = F(R_{\a}, \dot{R}_{\a}, a_{\a}),
\ee
where $\dot{R}_{\a} \equiv dR_{\a}/du_{\a}$.
In terms of the velocity field $V(u, r)$,
the equation above is equivalent to
\be
DV(u, r) = F(r, V(u, r), a(u, r)),
\label{DV}
\ee
where the derivative $D$ is defined by
\be
Df(u, r) \equiv e^{-\psi(u, r)} \frac{\del}{\del u}f(u, r) + V(u, r)\frac{\del}{\del r}f(u, r).
\label{comoving-derivative}
\ee
It is the time derivative of $u_{\a}$ along the trajectories of the thin shells.

In general,
if a physical quantity $f_{\a}(u_{\a})$ defined for each thin shell 
is to be replaced by a field $f(u, r)$ through the correspondence
$f(u, R_{\a}(u_{\a}(u))) = f_{\a}(u_{\a}(u))$ (for all $\alpha$),
we have
\be
\frac{df_{\a}}{du_{\a}} = 
\frac{du}{du_{\a}}\frac{\del f}{\del u}(u, R_{\a}) 
+ \frac{dR_{\a}}{du_{\a}}\frac{\del f}{\del r}(u, R_{\a}),
\ee
where $dR_{\a}/du_{\a}$ can be replaced by $V(u, R_{\a})$,
and $du/du_{\a}$ by $e^{-\psi}$.
Hence we are justified to use the replacement
\be
\frac{d}{du_{\a}}
\quad \rightarrow \quad
D \equiv e^{-\psi(u, r)} \frac{\del}{\del u} + V(u, r) \frac{\del}{\del r}
\label{d2D}
\ee
when a quantity $f_{\a}(u_{\a})$ is replaced by the corresponding space-time field $f(u, r)$.

The evolution of the Schwarzschild radius $a_{\a}$ for each layer is 
determined by the energy flux of pre-Hawking radiation,
and its evolution equation is \cite{Kawai:2013mda}
(See Ref.\cite{Kawai:2013mda} and Appendix B.)
\be
\frac{da_{\a}}{du_{\a}} = - \frac{N_0 \ell_p^2}{4\pi}\{u_{\a}, U\},
\label{dot-a-a}
\ee
where $U$ is the light-cone parameter in the infinite past
\footnote{
$U$ is the Eddington advanced time of the infinite past,
which can be used to label ingoing light-rays.
Ingoing light rays from the infinite past
are reflected at the origin 
and then turned into outgoing light rays,
which can be labelled by the Eddington retarded time $u_{\a}$
when it crosses the infalling thin shell labelled by $\a$.
There is thus a correspondence between $U$ and $u_{\a}$.
When the collapsing sphere is a shell,
$U$ can be identified with the Eddington retarded time 
for the Minkowski space 
enclosed within the inner surface of the shell.
}
and $\{\cdot, \cdot\}$ is the Schwarzian derivative defined by
\be
\{u_{\a}, f\} \equiv \left(\frac{\ddot{f}}{\dot{f}}\right)^2 - \frac{2\dddot{f}}{3\dot{f}}
\label{Schwarzian}
\ee
with the dots referring to derivatives with respect to $u_{\a}$.
Since
\be
\frac{dU}{du_{\a}} = e^{\tilde\psi_{\a}},
\label{tildepsia}
\qquad
\mbox{where} \quad \tilde\psi_{\a} \equiv \psi(u, 0) - \psi(u, R_{\a}),
\ee
we have
\be
\frac{da_{\a}(u_{\a})}{du_{\a}} =  - \frac{N_0 \ell_p^2}{4\pi}\{u_{\a}, U\} 
= - \frac{N_0 \ell_p^2}{12\pi} ( \dot{\tilde\psi}_{\a}^2 - 2\ddot{\tilde\psi}_{\a} ).
\label{daadua}
\ee

A special case of this equation is
\be
\frac{da_0(u)}{du} 
= - \frac{N_0 \ell_p^2}{12\pi} ( \dot{\psi}_0^2 - 2\ddot{\psi}_0 ),
\label{da0du0}
\ee
where 
\be
\psi_0 \equiv \psi(u, 0) = -\frac{1}{2}\int_0^{R_0(u)} dr' \; \frac{a'(u, r')}{r' - a(u, r') + r' V(u, r')}.
\label{psi0}
\ee
Eq.(\ref{da0du0}) determines the evolution of the Schwarzschild radius $a_0$ 
of the collapsing sphere.
Analogous to eq.(\ref{a-trajectory}),
it approximately gives
\be
\frac{da_{\a}(u_{\a})}{du_{\a}} \simeq - \frac{\sigma}{a^2_{\a}(u_{\a})}
\label{dadu}
\ee
for large $a_{\a}(u_{\a})$
in a collapsing sphere with a smooth mass distribution.

In terms of $a(u, r)$ and $\psi(u, r)$,
eq.(\ref{daadua}) is equivalent to
\be
Da(u, r) = - \frac{N_0 \ell_p^2}{12\pi} \left[ (D\tilde\psi(u, r))^2 - 2 D^2\tilde\psi(u, r) \right],
\label{Da}
\ee
where 
\be
\tilde{\psi}(u, r) \equiv \psi(u, 0) - \psi(u, r).
\ee

Given initial conditions on $\psi(u, r)$, $V(u, r)$ and $a(u, r)$,
eqs.(\ref{red-shift-general}), (\ref{DV}) and (\ref{Da})
determine the time evolution of the three functions $\psi(u, r)$, $V(u, r)$ and $a(u, r)$,
and they fix the geometry inside the collapsing sphere.

\section{Geometry of Collapsing Sphere}
\label{Geometry}

In the above,
we have decomposed a collapsing sphere into infinitely many infinitesimally thin collapsing shells
separated by infinitesimally thin layers of space filled with outgoing pre-Hawking radiation.
Assuming that the pre-Hawking radiation is dominated by massless dust,
the resulting metric in the continuous limit is expected to be the outgoing Vaidya metric 
modified by the red-shift factor $e^{\psi}$ (\ref{red-shift-general}):
\be
ds^2 = - e^{2\psi(u, r)} \left( 1 - \frac{a(u, r)}{r} \right) du^2 
- 2 e^{\psi(u, r)} du dr + r^2 d\Omega.
\label{metric}
\ee
Apart from the definition of $\psi(u, r)$ by eq.(\ref{red-shift-general}),
this is formally a totally generic spherically symmetric metric
since the spherical symmetry in 3+1 dimensions only allows
two parametric functional degrees of freedom.

There is an event horizon at $r = a^*$ if 
\be
\lim_{u\rightarrow\infty}a_0(u) = a^*.
\ee
There is no horizon if there is complete evaporation (\ref{CE}).

For the metric (\ref{metric}),
the light-like one-forms in the radial directions are
\bea
e^{\psi(u, r)} du
\quad \mbox{and} \quad
\xi \equiv e^{\psi(u, r)} \left( 1 - \frac{a(u, r)}{r} \right) du + 2 dr.
\label{xi}
\eea
The 1-form $e^{\psi(u, r)} du$ is interpreted as the differential 
of the Eddington retarded time $du_{\a}$ for $r = R_{\a}(u_{\a})$.
Constant-$u$ curves are outgoing light-like geodesics,
and $\xi$ vanishes on the ingoing like-like geodesics.

The function $m(u, r)$ defined by $m(u, r) = a(u, r)/2$
has an invariant geometric meaning \cite{mass} 
and can be interpreted as the mass inside 
the sphere of radius $r$ at the Eddington retarded time $u$.
The function $\rho(u, r)$ defined by
\be
\rho(u, r) \equiv \frac{1}{4\pi r^2} \frac{\del m(u, r)}{\del r}
\label{rho}
\ee
can be interpreted as the density function
of the collapsing matter sphere.

The physical degrees of freedom in the two functions $a(u, r)$ and $\psi(u, r)$
correspond to the freedom in the radial distribution of energy ($\rho(u, r)$),
and the freedom in the velocities of particles ($V(u, r)$).
For the special case of collapsing at the speed of light,
$\psi(u, r)$ is completely determined by $a(u, r)$.

According to their physical interpretation,
the parametric functions $a(u, r)$ and $\psi(u, r)$ 
are expected to satisfy the following conditions:
\bea
a(u, r) &=& a_0(u) 
\qquad \forall r \geq R_0(u),
\label{a0-cond}
\\
0 \; \leq \; a(u, r) &<& r
\qquad \forall r > 0,
\label{a-cond}
\\
a'(u, r) &\geq& 0
\qquad \forall r \geq 0,
\label{a'-cond}
\\
Da(u, r) &\leq& 0
\qquad \forall r \geq 0,
\label{Da-cond}
\eea
and
\bea
\psi(u, r) &=& 0
\qquad \forall r \geq R_0(u),
\label{psi0-cond}
\\
\psi'(u, r) &\geq& 0
\qquad \forall r \geq 0.
\label{psi'-cond}
\eea

The first constraint (\ref{a0-cond}) on $a(u, r)$
ensures that the geometry for $r \geq R_0(u)$ 
is given by that of the outgoing Vaidya metric (\ref{OGVM})
with the Schwarzschild radius $a_0(u)$.
The second constraint on $a(u, r)$ (\ref{a-cond})
makes sure that the Schwarzschild radius $a(u, r)$
is always hidden inside the thin shell at $r$,
and thus the factor $( 1 - a(u, r)/r )$ is always positive.
The third constraint on $a(u, r)$ (\ref{a'-cond}) corresponds to the statement 
that the energy density $\rho(u, r)$ (\ref{rho}) is non-negative.
The fourth constraint on $a(u, r)$ (\ref{Da-cond}) is
the assumption that pre-Hawking radiation
decreases the energy of each matter shell
in the collapsing sphere.
If $Da(u, r) = 0$ ($\forall \, r, u$),
the metric (\ref{metric}) describes a collapsing sphere of dust without radiation
and it should be equivalent to the Lama\^{i}tre-Tolman-Bondi metric.

The first constraint on $\psi$ (\ref{psi0-cond}) refers to the definition of $R_0$
that there is no collapsing matter for $r > R_0$.
The second constraint on $\psi$ (\ref{psi'-cond}) corresponds to the statement
that the red-shift factor is larger at smaller $r$.
Combining these two constraints on $\psi$,
we get
\be
\psi(u, r) \leq 0,
\ee
which means that the Eddington retarded time coordinate of an inner shell
suffers a red-shift factor relative to the time coordinate $u$ at distance.

\subsection{Energy-Momentum Tensor}
\label{EMT}

The Einstein equation determines
the energy-momentum tensor for the metric (\ref{metric}) to be
\bea
T_{uu} &=& 
\frac{1}{8\pi G} \frac{1}{r^2} e^{\psi} \left[ - \dot{a} + e^{\psi} \left( 1 - \frac{a}{r} \right) a' \right],
\label{Tuu}
\\
T_{ur} &=&
\frac{1}{8\pi G} \frac{1}{r^2} e^{\psi} a',
\label{Tur}
\\
T_{rr} &=& 
\frac{1}{8\pi G} \frac{2}{r} \psi',
\label{Trr}
\\
T_{\th\th} &=& 
\frac{1}{8\pi G}\left[
r\psi' + \frac{1}{2} a\psi' - \frac{3}{2} ra'\psi' + r^2 \left( 1 - \frac{a}{r} \right) (\psi')^2
- \frac{1}{2} r a'' + r^2 \left( 1 - \frac{a}{r} \right) \psi'' - e^{-\psi} r^2 \dot{\psi}' 
\right],
\nn
\\
\label{Tthth}
\\
T_{\phi\phi} &=&
\sin^2\th T_{\th\th}.
\label{Tphiphi}
\eea
The conservation of energy-momentum is ensured by the Einstein equation.
The weak energy condition demands that $T_{uu}$ and $T_{rr}$ be positive,
which are guaranteed by the requirements (\ref{a-cond})--(\ref{Da-cond}) and (\ref{psi'-cond}).

The energy-momentum tensor can be decomposed into three parts:
\be
T = T^{(out)} + T^{(in)} + T_{\th\th} d\Omega^2,
\label{T-decomp}
\ee
where $T \equiv T_{\mu\nu} dx^{\mu} \otimes dx^{\nu}$
is the total energy-momentum tensor,
$T^{(out)}$ represents the pre-Hawking radiation,
$T^{(in)}$ stands for the collapsing matter
and $T_{\th\th} d\Omega^2$ is the tangential component
induced by the collapse and radiation.
We have
\bea
T^{(out)} &=& T_{out} (e^{\psi} du \otimes e^{\psi} du),
\\
T^{(in)} &=& T_{in} \zeta \otimes \zeta,
\eea
where
\be
\zeta = k(u, r) du + 2dr
\label{zeta}
\ee
is supposed to be proportional to the velocity one-form of the collapsing matter.

The component form of the energy-momentum tensor (\ref{Tuu})--(\ref{Trr})
implies that
\bea
T_{out} &=&
- \frac{1}{8\pi G}
\frac{1}{r^2} \left[ e^{-\psi} \dot{a} - \frac{1}{2} \left( 1 - \frac{a}{r} \right) a' \right]
= - \frac{1}{8\pi G} \frac{1}{r^2} Da,
\\
T_{in} &=& \frac{1}{4} T_{rr} = \frac{1}{8\pi G} \frac{1}{2r} \psi',
\eea
and
\be
k(u, r) = 2 \frac{T_{ur}}{T_{rr}}.
\label{k}
\ee

As a confirmation of our interpretation about the decomposition (\ref{T-decomp}),
the expression of $T_{out}$ is precisely that of the outgoing energy flux 
for the outgoing Vaidya metric in an infinitesimal layer of space:
\be
T_{u_{\a}u_{\a}} = - \frac{1}{8\pi G} \frac{1}{r^2} \frac{da_{\a}}{du_{\a}},
\ee
with $da_{\a}/du_{\a}$ replaced by $Da$.
(See (\ref{comoving-derivative}) for the definition of the comoving derivative $D$.)

Using eq.(\ref{k}),
the velocity one-form $\zeta$ (\ref{zeta}) can be written as the superposition of
the two light-like one-forms (\ref{xi}) as
\be
\zeta = \frac{1}{2}\xi + \left[\frac{1}{2}\left(1-\frac{a}{r}\right)+V(u, r)\right] e^{\psi}du.
\label{zeta}
\ee
This is time-like if
\be
V(u, r) > - \frac{1}{2}\left(1-\frac{a}{r}\right),
\ee
which is precisely the condition
that the velocity $V(u, r)$ is smaller than the speed of light.
This is another confirmation of our interpretation of the decomposition (\ref{T-decomp}).

\subsection{Absence of Singularity at Surface of Collapse}
\label{NoSingularity}

As long as $a(u, r)$ and $\psi(u, r)$ are continuous functions at $r = R_0(u)$,
as the continuity conditions (\ref{a0-cond}) and (\ref{psi0-cond}) already imply,
there is no $\delta$-function contributions at the boundary
in the energy-momentum tensor components $T_{uu}, T_{ur}$ and $T_{rr}$.
To ensure the absence of $\delta$-function contribution in $T_{\th\th}$ and $T_{\phi\phi}$ at the boundary
(the absence of an additional membrane with non-zero tension at $r = R_0(u)$),
we need
\be
\left[ \frac{1}{2R_0} a'' - \left(1 - \frac{a_0}{R_0}\right)\psi'' + \dot{\psi}' \right]
\ee
to be free of $\delta$-function at $r = R_0(u)$.
This means that we need
\be
\left[ \frac{1}{2R_0} a' - \left(1 - \frac{a_0}{R_0}\right)\psi' + \dot{\psi} \right]
\label{check-delta}
\ee
to be continuous at $r = R_0(u)$.
(It is already known that 
$\left(1-\frac{a}{r}\right)$ is continuous at $r = R_0$.)

For $r > R_0$,
$a(u, r) = a_0(u)$ and so $a'(u, R_0^+) = 0$.
($R_0^+$ represents the limit of $r \rightarrow R_0$ from the side $r > R_0$.)
Similarly, $\psi(u, r) = 0$ for $r > R_0$,
and so $\psi'(u, R_0^+) = \dot{\psi}(u, R_0^+) = 0$.
Thus the expression (\ref{check-delta}) has a limit of $0$ at $r = R_0^+$.
For $r < R_0$,
using the general expression for the red-shift factor (\ref{red-shift-general}),
we have
\be
\psi'(u, R_0^-(u)) = \frac{1}{2}\frac{a'(u, R_0)}{R_0 - a_0 + a_0 \dot{R}_0},
\qquad
\dot{\psi}(u, R_0^-(u)) = - \frac{1}{2} \dot{R}_0 \frac{a'(u, R_0)}{R_0 - a_0 + a_0 \dot{R}_0},
\ee
which implies that (\ref{check-delta}) vanishes at $r = R_0^-$.
($R_0^-$ refers to the limit of $r \rightarrow R_0$ from the side of $r < R_0$.)
Hence the expression (\ref{check-delta}) is indeed continuous at $r = R_0$,
demonstrating that there is no $\delta$-function contribution in $T_{\th\th}$,
or in any component of the energy-momentum tensor.
We have thus proven that
the metric (\ref{metric}) with $\psi$ defined by eq.(\ref{red-shift-general})
does not suffer singularity at the interface $r = R_0$.

\subsection{Collapsing at Speed of Light}
\label{CollapseLight}

For the case in which both the collapsing matter 
and the pre-Hawking radiation are light-like,
the energy-momentum tensor (\ref{T-decomp}) is decomposable 
as the sum of the ingoing and outgoing light-like energy fluxes,
apart from the pressure $T_{\th\th}$ in the tangential directions.
Indeed, 
the velocity one-form is light-like
($\zeta = \xi/2$)
according to eq.(\ref{zeta})
when $V(u, r)$ is given by eq.(\ref{V-lightlike}),
and the energy-momentum tensor is decomposable into 
the three components:
\bea
T_{out} &\equiv& 
= - \frac{1}{8\pi G}
\frac{Da(u, r)}{r^2},
\label{Tout}
\\
T_{in} &\equiv& \frac{1}{8\pi G}
\frac{1}{2r^2}\frac{a'}{1 - \frac{a}{r}},
\label{Tin}
\\
T_{\th\th} &\equiv& \frac{1}{8\pi G}
\left[
\frac{1}{2}\frac{1}{r-a}(a+ra')a' + \frac{1}{2}ra''-e^{-\psi}r^2\frac{\del}{\del u}\left(\frac{a'}{r-a}\right)
\right]
\nn \\
&=&
e^{-2\psi} \frac{r^2}{r-a} \frac{\del}{\del r}\left[ e^{2\psi} r^2 T_{out} \right].
\label{Tthth}
\eea

According to eq.(\ref{red-shift}),
the red-shift factor is
\be
\psi(u, r) = - \Theta(R_0(u) - r) \int_r^{R_0(u)} dr' \; \frac{a'(u, r')}{r' - a(u, r')}.
\label{psi-KMY}
\ee
With $\psi(u, r)$ given by this expression,
the space-time metric is completely determined by $a(u, r)$.

For a sphere collapsing at the speed of light for a long period of time, 
the equation for ingoing null geodesics (\ref{R-alpha}) 
can be used to solve $R_{\a}(u_{\a})$ as a function of $a_{\a}(u_{\a})$ iteratively 
with the help of eq.(\ref{dot-a-a}),
so that $R_{\a}$ can be expressed as a function of $a_{\a}$.
First,
eq.(\ref{R-alpha}) implies that
\be
R_{\a}(u_{\a}) = a_{\a}(u_{\a}) - 2 R_{\a}(u_{\a}) \dot{R}_{\a}(u_{\a}).
\label{R-iterate}
\ee
If $R_{\a}$ has stayed close to $a_{\a}$ over an extremely long period of time,
to the lowest order approximation,
we have $R_{\a} \simeq a_{\a}$.

To have a better approximation,
one can use (\ref{R-iterate}) with the last term $2R_{\a}\dot{R}_{\a}$ on the right
replaced by its approximation $2a_{\a}\dot{a}_{\a}$ at the lowest order.
That is,
\be
R_{\a}(u_{\a}) 
\simeq 
a_{\a}(u_{\a}) - 2 a_{\a}(u_{\a})\dot{a}_{\a}(u_{\a}) 
\simeq 
a_{\a}(u_{\a}) + \frac{2\sigma}{a_{\a}(u_{\a})},
\label{delta-Ra}
\ee
where we used eq.(\ref{dadu}) for $\dot{a}_{\a}(u_{\a})$
as the lowest order approximation of eq.(\ref{dot-a-a}).
This is a better estimate of $R_{\a}(u_{\a})$,
and in particular we have
\be
\Delta R_0(u) \equiv R_0(u) - a_0(u) 
\simeq - 2 a_0 \dot{a}_0
\simeq \frac{2\sigma}{a_0(u)}.
\label{delta-r}
\ee

For an approximation to the next order,
we can use eq.(\ref{delta-r}) to estimate 
the second term on the right hand side in eq.(\ref{R-iterate})
to get an even better estimate of $R_{\a}(u_{\a})$.
This iteration can go on for ever,
leading to an all order expression for $R_{\a}(u_{\a})$ as a function of $a_{\a}(u_{\a})$ only.
The asymptotic form of $R_{\a}(u_{\a})$ depends on $u_{\a}$ only through $a_{\a}(u_{\a})$.

To the approximation of (\ref{delta-Ra}),
every matter shell stays at a coordinate difference
\be
\Delta R_{\a} \simeq \frac{2\sigma}{a_{\a}}
\label{delta-R-a}
\ee
outside its Schwarzschild radius $a_{\a}$.
It implies that
\be
r - a(u, r) \simeq \frac{2\sigma}{r},
\qquad
\mbox{or equivalently,}
\qquad
a(u, r) \simeq r - \frac{2\sigma}{r},
\label{r-a}
\ee
which can also be derived from eqs.(\ref{V-lightlike}) and (\ref{Da}).
In particular,
\be
a_0 \simeq R_0 - \frac{2\sigma}{R_0}.
\label{r0-a0}
\ee
Note that eq.(\ref{r-a}) is valid only for $r \gg \sqrt{2\sigma}$.

Using eq.(\ref{dadu}),
it is approximately
\bea
T_{out} &\simeq& \frac{1}{8\pi G} \frac{\sigma}{a^4},
\\
T_{in} &\simeq& \frac{1}{8\pi G} \frac{1}{4\sigma},
\label{Tin}
\\
T^{\th}{}_{\th} &\simeq& \frac{1}{8\pi G}\frac{1}{2\sigma}
\eea
for $r < R_0(u)$ but $r \sim R_0(u)$.
Only the outgoing radiation $T_{out}$ is present for $r > R_0(u)$.

\section{Asymptotic Approximation}
\label{Asymptotic}

In a gravitational collapse 
where the gravitational force dominates over other interactions,
each matter shell in the collapsing sphere 
is accelerated to approach to the speed of light.
The description of a collapsing sphere at the speed of light in Sec.\ref{CollapseLight}
is therefore a good approximation for a generic matter sphere
that has been going through gravitational collapse for a sufficiently long time,
until $a_{\a}$ becomes too small
for the approximation considered above to be valid.

During this long period of time when $R_{\a}$ stays extremely close to $a_{\a}$
and eq.(\ref{delta-R-a}) is a good approximation
for most of the matter shells in a collapsing sphere,
the collapsing sphere demonstrates a universal behavior,
and its geometry can be approximated by the same configuration for generic gravitational collapses.
This configuration is characterized by eqs.(\ref{r-a}) and (\ref{r0-a0}),
and we shall refer to it as the asymptotic black hole.

In the phase of the asymptotic black hole,
the matter sphere must be very massive
so that the pre-Hawking radiation is extremely weak,
hence the configuration looks almost static for a distant observer.
The surface of the collapsing sphere roughly stays at a constant distance
outside the Schwarzschild radius $a_0(u)$,
and the asymptotic black hole appears to be 
parametrized only by the Schwarzschild radius $a_0(u)$ of the whole sphere,
as the geometry changes adiabatically with the change in $a_0(u)$.

Let us now give more details about the asymptotic black hole.
Using eq.(\ref{r-a}),
the red-shift factor (\ref{psi-KMY}) of an asymptotic black hole is
\bea
\psi(u, r) &\simeq& - \Theta(R_0(u) - r) \int_r^{R_0(u)} dr \frac{a a'}{2\sigma}
\nn \\
&=& - \Theta(R_0(u) - r) \frac{a_0^2 - (r - 2\sigma/r)^2}{4\sigma}
\nn \\
&\simeq& - \Theta(R_0(u) - r) \frac{R_0^2 - r^2}{4\sigma} + {\cal O}\left(\frac{\sigma}{R_0^2}\right),
\label{psi-asymp}
\eea
where terms of order $\log(a_0)$ or higher are ignored.

For $r$ sufficiently close to $a_0$,
i.e. $| a_0 - r | \ll a_0$,
it gives
$\psi(u, r) \simeq - \Theta(R_0(u) - r) \frac{R_0(R_0 - r)}{2\sigma}$.
This means that $\exp(\psi(u, r))$ goes to zero quickly 
over a change in $r$ of order $\sigma/R_0$ below the surface of the collapsing sphere.
We can approximate the red-shift factor by
\be
e^{\psi(u, r)} \simeq 
e^{-\Theta(R_0 - r)\frac{R_0^2 - r^2}{4\sigma}}
\simeq
\left\{
\begin{array}{lll}
1 & \mbox{for} & r \geq R_0(u),
\\
e^{- \frac{R_0(R_0 - r)}{2\sigma}} & \mbox{for} & r < R_0, \quad (r \sim R_0)
\\
e^{-\frac{R_0^2}{4\sigma}} \sim 0 & \mbox{for} & r \ll R_0.
\end{array}
\right.
\label{red-shift-aBH}
\ee
This red-shift factor will help us answer many questions below.

\subsection{Metric for Asymptotic Black Hole}

With the red-shift factor given by eq.(\ref{red-shift-aBH}),
the metric (\ref{metric}) is approximately given by
\be
ds^2 \simeq
\left\{
\begin{array}{ll}
- \left(1 - \frac{a_0}{r}\right) du^2 - 2 du dr + r^2 d\Omega^2
& r \geq R_0(u),
\\
- e^{-\frac{R_0(R_0 - r)}{\sigma}} \left( \frac{2\sigma}{r^2} \right) du^2 
- 2 e^{-\frac{R_0(R_0 - r)}{2\sigma}} du dr + r^2 d\Omega
& r \leq R_0(u) \quad (r \sim R_0(u)),
\end{array}
\right.
\label{metric-collapse-0}
\ee
where $a_0(u) = R_0(u) - 2\sigma/R_0(u)$.
We are ignoring the region where $R_0 - r \gg \sigma/a_0$,
which is relatively frozen.
An important feature of this metric is that
the time dependence of the configuration enters only through $R_0(u)$.

Note that
factors of the form $(r^2/\sigma)^n$ in the first two terms 
are of higher orders in comparison with the exponential factors
in the large $R_0$ expansion.
However,
by tuning the exponent to vanish exactly at $r = R_0$,
factors of the form $(r^2/\sigma)^n$ is fixed 
at the leading order by the continuity of the metric at $r = R_0$.

The energy-momentum tensor for the asymptotic black hole
can be obtained straightforwardly from the results in Secs.\ref{EMT} and \ref{CollapseLight}.
In addition to the pre-Hawking radiation ($T_{out}$)
and an ingoing collapsing matter sphere ($T_{in}$),
the energy-momentum tensor includes another term $T_{\th\th}d\Omega^2$
corresponding to a very large pressure in the tangential directions of the collapsing shells
induced by the pre-Hawking radiation ($T^{\th}{}_{\th}$) \cite{Kawai:2015uya}.
(See eq.(\ref{Tthth}).)
The weak energy condition is satisfied
but the dominant energy condition is violated \cite{Kawai:2015uya}.
(Note that the violation of the dominant energy condition
is not unexpected in phenomena involving quantum tunnelling.)

For a distant observer,
a large collapsing sphere looks almost static,
and to the approximation that $a(u, r) = a(r)$ is time-independent,
the metric (\ref{metric-collapse-0}) can be presented 
in terms of the time coordinate analogous to the coordinate $t$ in Schwarzschild solution.
It is defined by
\be
dt = \left\{
\begin{array}{ll}
du + \frac{dr}{1 - \frac{a}{r}}
& (r \geq R_0),
\\
du + e^{\frac{R_0(R_0-r)}{2\sigma}}\frac{r^2}{2\sigma} dr
& (r < R_0).
\end{array}
\right.
\ee
Eq.(\ref{r0-a0}) ensures the continuity of the definition of $t$ at $r = R_0$.
In terms of the new coordinate $t$,
the metric is
\be
ds^2 \simeq
\left\{
\begin{array}{ll}
- \left(1 - \frac{a_0}{r}\right) dt^2 + \frac{1}{1 - \frac{a_0}{r}} dr^2 + r^2 d\Omega^2
& (r \geq R_0),
\\
- e^{-\frac{R_0(R_0 - r)}{\sigma}} \left( \frac{2\sigma}{r^2} \right) dt^2 
+ \frac{r^2}{2\sigma} dr^2 + r^2 d\Omega
& (r \leq R_0).
\end{array}
\right.
\label{metric-equilibrium}
\ee
As $r \rightarrow \infty$,
$t$ is the time coordinate of the Minkowski space at infinity.
With $R_0$ set to be a constant,
this metric can also be viewed as an approximation 
of the metric of an asymptotic black hole in equilibrium
with the environment at the Hawking temperature.

A very important lesson we can learn from 
either the metric (\ref{metric-collapse-0}) or (\ref{metric-equilibrium}) is that
there is a huge red-shift factor that practically freezes up 
everything that is beyond a coordinate difference of order $\sigma/a_0$
below the surface of the collapsing sphere.
To a distant observer,
all dynamical degrees of freedom reside
on the surface layer of the collapsing sphere
of physical thickness of order
\be
\sqrt{g_{rr}} \frac{2\sigma}{a_0}
\simeq \left(1-\frac{a_0}{R_0}\right)^{-1/2}\frac{2\sigma}{a_0} \simeq \sqrt{2\sigma}.
\ee
Everything beneath this layer is frozen.
There is no horizon,
but it would appear to a distant observer that
particles falling through the surface layer 
can never come back due to the large red shift.

Notice that it is $\sqrt{2\sigma}$
that is the characteristic length scale in the geometry of the asymptotic black hole.
It can in principle be much longer than the Planck length
if the number $N_0$ of massless fields is large.
For large $N_0$,
the geometry does not involve Planck-scale physics anywhere \cite{Kawai:2015uya}.
This model offers an opportunity to resolve the information loss paradox
without resorting to Planck-scale physics.

More general metrics than eq.(\ref{metric-collapse-0}) 
were given in Ref.\cite{Kawai:2014afa} for the late stage of gravitational collapse.
(See Appendix D.)

\subsection{Pre-Hawking Radiation}
\label{HawkingRadiation}

As we have mentioned,
a crucial feature that distinguishes the KMY model 
from the conventional models is that
pre-Hawking radiation contributes to the energy-momentum tensor in Einstein's equation
not only at distance,
but also at the neighborhood of the collapsing matter.
This contribution can be computed as the expectation value 
of the energy-momentum operator \cite{Kawai:2013mda}.

The total energy flux of the pre-Hawking radiation
is determined by $\psi_0$ (\ref{psi0}).
According to eq.(\ref{red-shift-aBH}),
it is approximately
\be
e^{\psi_0} \equiv \frac{dU}{du} \simeq e^{- \frac{R_0^2}{4\sigma}}
\label{huge-red-shift}
\ee
for an asymptotic black hole.
Hence
\be
\dot{\psi}_0 \simeq - \dot{R}_0 \frac{R_0}{2\sigma}
\simeq \frac{1}{2R_0},
\ee
and the evolution equation for $a_0(u)$ (\ref{da0du0}) becomes
\be
\frac{da_0(u)}{du}
= - \frac{N_0 \ell_p^2}{12\pi} ( \dot{\psi}_0^2 - 2\ddot{\psi}_0 )
\simeq - \frac{N_0 \ell_p^2}{48\pi} \frac{1}{R_0^2}
\simeq - \frac{\sigma}{a_0^2},
\label{dadu-aBH}
\ee
for $R_0^2 \gg \sigma$.
This is in agreement with eq.(\ref{a-trajectory}),
which is the usual expression for Hawking radiation.
It implies complete evaporation (\ref{a-u-1/3})
at the macroscopic scale.

The spectrum of pre-Hawking radiation can be obtained by
computing the Bogoliubov transformation induced by
the time-independent background of the collapsing sphere.
It is approximately the radiation spectrum at the Hawking temperature.
For more details, see Appendix B.

\subsection{Point-Particle Trajectory}
\label{PointParticle}

In this section,
we consider the motion of a free particle 
as a probe of the geometry inside a matter sphere 
in gravitational collapse.
For the purpose of illustration,
it is sufficient to consider the approximate form of the metric 
eq.(\ref{metric-equilibrium}). 

To simplify the notation,
we write the metric (\ref{metric-equilibrium}) as
\be
ds^2 = - e^{2\psi} B dt^2 + \frac{1}{B} dr^2 + r^2 d\Omega^2,
\ee
where $\psi$ is given by eq.(\ref{red-shift-aBH})
and $B \equiv \frac{2\sigma}{r^2}$ for $r < R_0$.

The action for a point particle of mass $m$ is
\be
S = - m \int dt \; \sqrt{ e^{2\psi} B - \frac{1}{B} \dot{r}^2 },
\ee
where we have omitted the angular degrees of freedom
because we will focus on motions in the radial direction.
($\dot{r}$ refers to $dr/dt$.)
The conjugate momentum of $r$ 
and the Hamiltonian are
\be
p_r = \frac{m}{\sqrt{e^{2\psi} B - \frac{1}{B} \dot{r}^2}} \frac{\dot{r}}{B}
\qquad
\mbox{and}
\qquad
H = \frac{m e^{2\psi} B}{\sqrt{e^{2\psi} B - \frac{1}{B} \dot{r}^2}}.
\ee
For a particle with a given energy $H = E$,
we have
\be
\sqrt{e^{2\psi} B - \frac{1}{B} \dot{r}^2} = \frac{m}{E}e^{2\psi} B > 0.
\ee
There is thus an inequality
\be
e^{2\psi} B - \frac{1}{B} \dot{r}^2 > 0,
\ee
or equivalently,
\be
|\dot{r}| < B e^{\psi}.
\ee

The time $T$ for a distant observer to see a particle travelling 
from $r = r_0 < R_0$ to $R_0$ (or the other way) is thus
\be
T \simeq \int_{r_0}^{R_0} \frac{dr}{|\dot{r}|} 
> \int_{r_0}^{R_0} dr \; \frac{r_0^2}{\sigma}e^{\frac{R_0(R_0-r)}{2\sigma}}
\simeq r_0 e^{\frac{R_0(R_0-r_0)}{2\sigma}},
\ee
assuming that $R_0(R_0 - r_0)/\sigma \gg 1$.
Consider, for example,
$r_0 = R_0 - 100 \sigma/a_0$.
The expression above gives roughly
\be
T \geq 10^{22} \times R_0.
\ee
This is an extremely long time 
to fall through an extremely short distance.
For $R_0$ of the order of the sun's radius (about $5$ light seconds),
$T$ is 30000 times the age of the universe ($10^{10}$ years).
Although this is still much shorter than the time 
it takes to evaporate a sun's mass by Hawking radiation ($10^{67}$ years),
this is the same as saying that,
for most practical purposes 
from the viewpoint of a distant observer,
the particle will never be able get anywhere deeper than,
say, $100$ times the Planck length under the surface of the collapsing sphere
(assuming that $\sqrt{\sigma/2}$ is of the order of the Planck length).

From the viewpoint of a distant observer
this is practically the same as having a ``horizon'' roughly at $a_0$
and everything behind this ``horizon'' can never come out.
On the other hand,
according to the mathematical definition of apparent horizon or event horizon,
there is no horizon in space-time.

\subsection{Absence of Singularity at Origin}
\label{r=0singularity}

While it has been proven in Sec.\ref{NoSingularity}
that there is no singularity at the surface of the collapsing sphere ($r = R_0(u)$),
we show here that it is possible to avoid the singularity at $r = 0$ in the asymptotic black hole
and comment on generic gravitational collapses.

In the above,
we have used eq.(\ref{r-a}) inside the asymptotic black hole.
But eq.(\ref{r-a}) is valid only for $r \gg \sqrt{2\sigma}$.
We have not yet discussed the geometry for $r < \sqrt{2\sigma}$.
In this subsection,
we shall assume that $N_0 \gg 1$
such that $\sqrt{2\sigma} \gg \ell_p$
and study the geometry of an asymptotic black hole near the origin.

To understand the geometry near the origin,
we first note that there is a huge red-shift factor (\ref{huge-red-shift})
between the time coordinate $u$ outside the collapsing sphere
and the time coordinate $U$ near the origin.
As this number is exponentially larger than the time scale for evaporation
(it is $10^{10^{86}}$ for $R_0$ of order of the sun's radius
and $\sqrt{\sigma}$ of order of the Planck scale.),
we are justified to assume that
$a(u, r)$ is nearly independent of $u$ for $r \sim 0$.
That is,
\be
\dot{a}(u, r) \simeq 0
\qquad
\mbox{for} 
\quad 
r \sim 0.
\label{dadu=0}
\ee
Together with eq.(\ref{r-a}),
we have $a(u, r) \simeq a(r)$ \; $\forall r \leq R_0$.

While $V(u, r)$ is approximately given by eq.(\ref{V-lightlike})
for an asymptotic black hole,
$a(r)$ can be fixed as follows.
Using eqs.(\ref{red-shift}) and (\ref{tildepsia}), 
we find
\be
\tilde{\psi}_{\a} \simeq - \int_0^{R_{\a}} dr \, \frac{a'(r)}{r-a(r)}.
\ee
One can then check that,
using eq.(\ref{dadu=0})
in the limit $r \rightarrow 0$,
the evolution equation of $a(u, r)$ (\ref{Da}) can be solved 
as an expansion in $r$:
\be
a(r) \simeq A r^2 + {\cal O}(r^3)
\ee
for an arbitrary positive constant $A$.
As a result,
$(1-a(r)/r)$ approaches to a constant in the limit $r \rightarrow 0$.
It follows that all components of the energy-momentum tensor 
(\ref{Tuu})--(\ref{Tphiphi}) approach to zero as $r \rightarrow 0$,
and the metric (\ref{metric-collapse-0}) approaches to the Minkowski space,
without singularity.

Strictly speaking,
the calculation above only proves that 
there exist asymptotic black holes 
(which are stablized for $r \leq R_0$ in the sense of eqs.(\ref{r0-a0}) and (\ref{dadu=0}))
that are regular at $r = 0$.
We can never rule out the possibility that
singularity exists simply due to an improper choice of initial condition
or the time evolution equation (\ref{DV}).
A more interesting task is to characterize the singularity
generated by a gravitational collapse
with a regular initial condition,
and compare it with the singularity in a conventional model.

\begin{figure}
\vskip-2em
\center
\includegraphics[scale=0.45,bb=0 100 500 450]{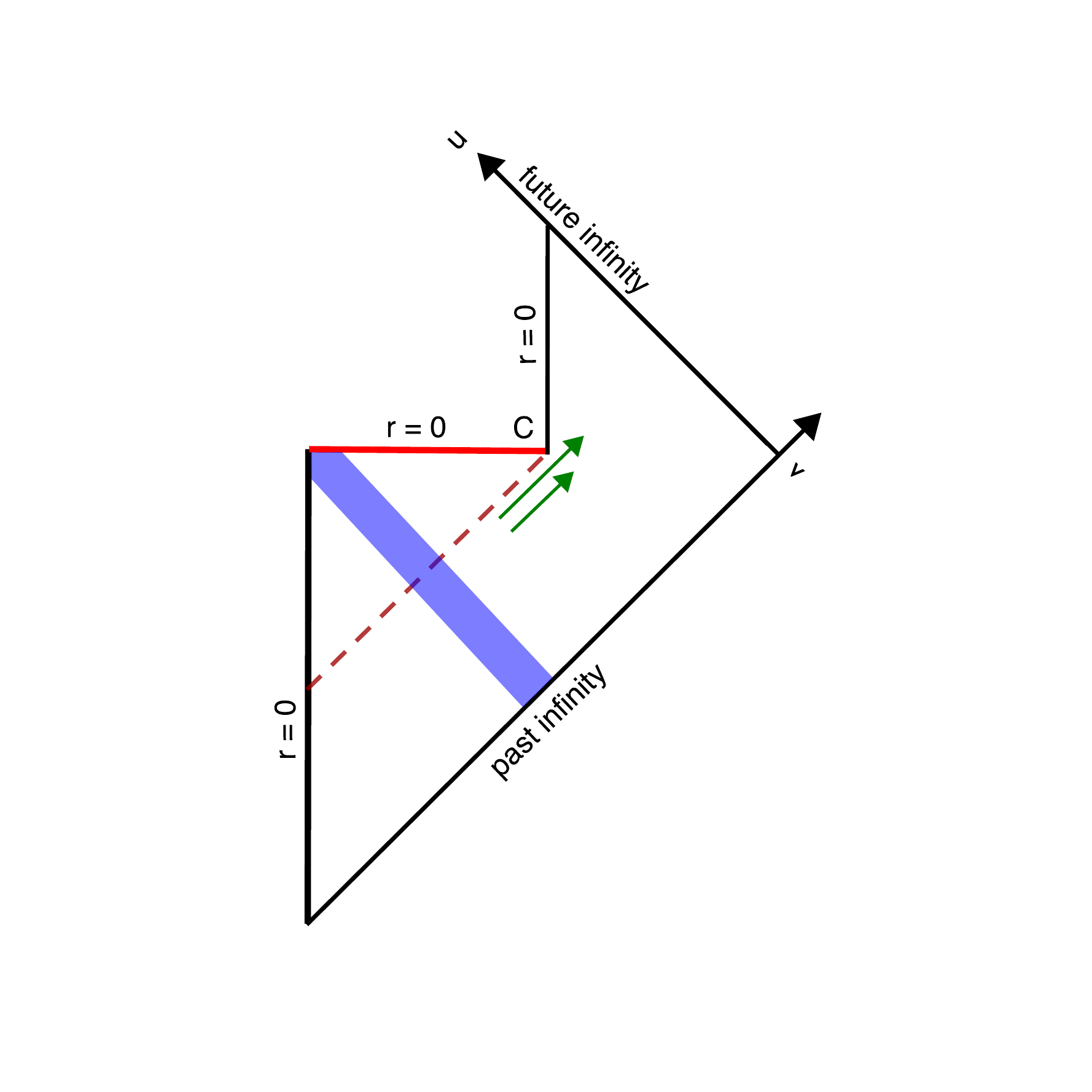}
\includegraphics[scale=0.4,bb=0 100 500 450]{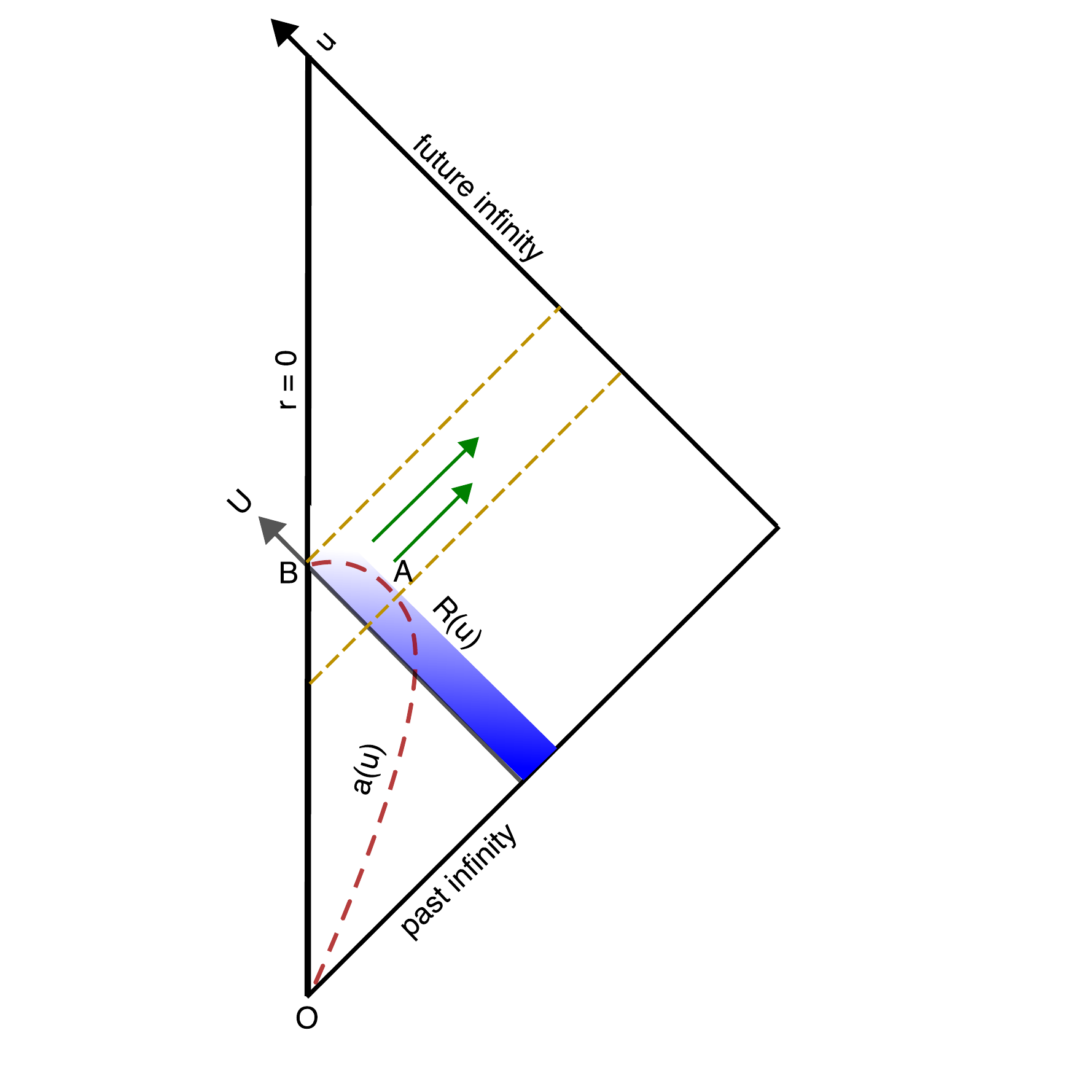}
\vskip2em
(a)\hskip17em(b)
\caption{\small
(a) Penrose diagram for the conventional model:
For a distant observer,
the space-like singularity at $r = 0$ can be seen only at point $C$
when the black hole is completely evaporated.
(b) Penrose diagram for the KMY model:
The red dash curve represents the fictitious Schwarzschild radius defined by 
the metric outside the collapsing spherical shell (blue strip) through analytic continuation.
For a distant observer,
the time-like singularity at $r = 0$ can be seen only very briefly around point $B$,
where the Schwarzschild radius is approaching to zero.
The yellow dash lines mark roughly the period of time when 
pre-Hawking radiation (green arrows) cannot be ignored.
}
\label{Penrose-KMY}
\vskip1em
\end{figure}

In a conventional model,
there is a space-like singularity at $r = 0$ inside the horizon.
This is a signal of the failure of low-energy physics.
However, 
for a distant observer, 
the singularity is visible only at the instant of complete evaporation.
Quantum gravity is needed only for a small space-time region
presumably of Planck scale.
(See Fig.\ref{Penrose-KMY}(a) \cite{Hawking:1974sw}.)

For the KMY model,
if the singularity at $r = 0$ exists,
it is a time-like (naked) singularity
far away from the surface of the collapsing sphere
until the last moment of evaporation.
It is irrelevant to the information loss paradox 
--- about how information is passed 
from the collapsing matter to pre-Hawking radiation,
which happens mainly near the surface,
except the last bit of evaproation.

For a distant observer,
the huge red-shift factor (\ref{huge-red-shift}) implies that, 
if there is no singularity at $r = 0$ in the initial state,
and it takes some time $\Delta U = \epsilon$ 
for the singularity to emerge at $r = 0$ out of the infalling matter,
the time it takes for a distant observer would be 
\be
\Delta T = \epsilon e^{\frac{R_0^2}{4\sigma}}.
\ee
This means that even if $\epsilon$ is of Planck scale,
$\Delta T$ is about $10^{10^{86}}$ years
if $R_0$ is about the size of the sun
and $\sqrt{\sigma}$ about the Planck time.
Of course, 
$\Delta T$ should not be larger than 
the time it takes for the whole star to evaporate.
What we learn from the calculation above is that
unless there is a singularity at the origin in the beginning,
from the viewpoint of a distant observer,
the origin can develop a singularity only 
briefly before the instant of complete evaporation.
Therefore,
just like in Sec.\ref{PointParticle},
we have seen how the huge red-shift factor (\ref{huge-red-shift})
serves as an effective horizon for the asymptotic black hole
for a point particle,
this red-shift factor also acts like a horizon to 
protect the singularity at the origin from being seen by a distant observer
until the last moment of evaporation
(see Fig.\ref{Penrose-KMY}(b)).
This feature is very similar to the story of the conventional model.

\section{Misconceptions}
\label{Misconceptions}

The explicit solution of the metric for a collapsing sphere including pre-Hawking radiation 
allows us to prove or disprove claims about black holes through explicit calculations,
without ignoring the back-reaction of the pre-Hawking radiation,
and therefore to clarify a few misconceptions.

\Misconception{1}{
If a gravitational collapse leads to the formation of an apparent or event horizon,
the horizon will still form if the gravitational collapse
is modified by an extremely weak outgoing radiation.
}

This erroneous statement is often used to justify 
the assumption of a black-hole apparent horizon 
in a gravitational collapse 
by ignoring the back-reaction of the Hawking radiation.
However,
as it was shown in Refs. \cite{Kawai:2013mda,Ho:2015fja,Ho:2015vga},
while the gravitational collapse of massless dust in classical gravity
will certainly lead to a black-hole horizon,
turning on an arbitrarily weak outgoing radiation
will obstruct the formation of any horizon,
as long as the radiation leads to a complete evaporation.
(Please refer to Refs.\cite{Ho:2015fja,Ho:2015vga} for a comprehensive explanation.)
Note that, 
as long as the radiation is finite 
(not approaching to zero,
i.e. $|\dot{a}| > \epsilon$ for any given constant $\epsilon > 0$),
the energy of an arbitrarily large star can always be exhausted within a finite time
(shorter than $2M/\epsilon$ for a star of mass $M$).
If the pre-Hawking radiation does not lead to complete evaporation,
the metric (\ref{metric}) would imply
the existence of a horizon.

The origin of the problem in this misconception 
can be related to the fact that
the Schwarzschild solution is degenerate
as it describes two physically different states:
the black hole and the white hole.
Consider a star whose radius $R_0$
is slightly larger than the Schwarzschild radius $a_0$ 
but the difference $R_0-a_0$ is too small to be detected by a distant observer.
If there is no radiation detected by the distant observer,
he is inclined to assume that it is a black hole.
However,
in principle it can be a star with an outgoing radiation
that is too weak to be observed,
and it is possible that the star will eventually evaporate completely,
without ever creating an apparent horizon.
(See Appendix A.)

It is perhaps still intuitively puzzling from the perspective of an infalling observer,
who falls in with the collapsing sphere.
How is it possible for him to see the whole star evaporate
before the horizon forms,
due to sparse creation of quantum particles?
More specifically,
an infalling observer falls to the origin within a proper time of order $M$
($M$ is the initial mass of the collapsing star),
while the power of pre-Hawking radiation is only of order $1/M^2$,
how can the whole star evaporate before
the observer reaches the origin?
This question can be answered quantitatively via calculations,
which is shown in Appendix C.
Roughly speaking,
the infalling observer in free fall is accelerated to 
approach to the speed of light
so that his proper time runs much slower,
while the intensity of pre-Hawking radiation gets much stronger,
compared with an observer at a constant distance from the star.

\Misconception{2}{
Hawking radiation (or Hawking-like radiation
--- including pre-Hawking radiation)
can appear only if 
there is (or will be) a black-hole horizon.
}

An intuitive interpretation of Hawking radiation is
to imagine a virtual pair of particles created near the horizon,
with one particle going to distance and the other falling into the horizon.
However, 
this does not mean that it is impossible to 
have Hawking-like radiation created under different circumstances.
The quantum field theory explanation of Hawking radiation is a nontrivial Bogoliubov transformation.
There have been a series of works \cite{HR1,HR2}
about Hawking-like radiation due to nontrivial evolutions of space-time.
The conclusion established by these works
is that it is possible to have Hawking-like radiation 
even when there is no (apparent or event) horizon.
What is really needed is an exponential relation between 
the parametric light-like coordinate $U$ for 
the incoming light rays from the infinite past 
and the light-like coordinate $u$ for the outgoing light rays to the infinite future.
(See Appendix B.)
It turns out that
the exponential relation between $U$ and $u$ for the KMY model
differs by a minus sign in the exponent from the conventional models
of gravitational collapse,
and as a result the Bogoliubov transformation coefficients 
are different by complex conjugation.
Nevertheless,
the spectrum of pre-Hawking radiation remains the same
\cite{Kawai:2013mda}.

\Misconception{3}{
If pre-Hawking radiation contributes to the energy-momentum tensor
at the surface of the collapsing sphere,
there would be a huge energy flux at the surface of collapse 
as it approaches to the Schwarzschild radius.
}

This statement has been used to argue why 
the space around the horizon
has to be a vacuum state
for an observer in free fall \cite{Unruh:1977ga},
and why Hawking radiation exists as classical radiation 
only at distance in conventional models.
However,
the blue-shift factor does not diverge if there is no horizon.

In the absence of the horizon,
one might still wonder if the pre-Hawking radiation would be too strong
at the surface of the collapsing sphere to use low energy physics.
This was checked in Ref.\cite{Ho:2015vga},
and it was checked that pre-Hawking radiation is extremely weak
everywhere.
The mechanism behind the calculation is the following.
The larger the pre-Hawking radiation is,
the faster the Schwarzschild radius shrinks,
and thus a larger separation between $R_0$ and $a_0$,
implying a smaller blue-shift factor.
As the energy flux density of the pre-Hawking radiation at distance is proportional to $1/a_0^4$,
and the blue-shift factor at $R_0$ is proportional to $a_0^2$,
the energy flux density of the pre-Hawking radiation at $R_0$
is proportional to $1/a_0^2$.
The energy flux density is thus extremely weak 
at the surface of the collapsing sphere for large $a_0$.
Hence,
even though we have not assumed the vacuum state for an observer in free fall
near the surface of the collapsing sphere,
for an asymptotic black hole with an astronomical mass,
the radiation flux is so weak that it would be hard to be distinguished from the vacuum state.

\section{Comments}
\label{Comments}

\subsection{Information Loss Paradox}

In conventional models of black holes \cite{Hawking:1974sw},
the gravitational collapse of a star can lead to an apparent horizon.
Hawking radiation is created outside the horizon,
and eventually the whole star evaporates.
This picture leads to a conflict between locality and unitarity in low energy effective theories.
Unitarity demands that Hawking radiation 
carries the full information of the collapsing matter,
but locality forbids the matter deep inside the horizon
to transfer its information efficiently to the Hawking radiation outside the horizon
\cite{Mathur:2009hf}.
This is the information loss paradox \cite{Hawking:1974sw,Mathur:2009hf}.

In the KMY model,
there is no horizon unless it already exists in the initial state.
No information available to a distant observer 
becomes hidden behind a horizon as a result of gravitational collapse.
In fact,
even if the formula of pre-Hawking radiation is modified 
such that a remnant black hole survives with a horizon,
as the pre-Hawking radiation is always created at the surface of the collapsing matter,
there is no need to sacrifice locality for unitarity
\cite{Kawai:2015uya,Ho:2015vga}.
This approach to solving the information loss paradox was also proposed in 
Refs. \cite{Vachaspati:2006ki,Saini:2015dea,Mersini-Houghton:2014zka,Mersini-Houghton:2014cta}.

\subsection{Black-Hole Entropy and Brick Wall Model}

As the spectrum of pre-Hawking radiation in the KMY model
is approximately the same
as that in the conventional model (\ref{a-trajectory}),
we arrive at the same expression
\be
S = \frac{\mbox{Area}}{4} = \pi a_0^2
\label{S}
\ee
for the entropy of an asymptotic black hole
based on the thermodynamical formula
\be
dS = dU/T_H,
\ee
where $T_H$ is taken to be the (pre-)Hawking temperature
$T_H = 1/(4\pi a_0)$.

In Sec.\ref{Asymptotic},
we found that,
with respect to a distant observer,
everything happening inside the collapsing sphere looks frozen,
except that the outermost shell of thickness $\Delta R_0$ (\ref{delta-r})
at the surface remains exponentially comparatively more active.
This result not only motivates a new membrane paradigm,
it also gives an intuitive explanation of 
the area law of the Bekenstein-Hawking entropy (\ref{S}),
apart from the numerical factor of $1/4$,

The KMY model also gives a new interpretation 
and some modification to
't Hooft's brick wall model \cite{'tHooft:1984re,'tHooft:1996tq},
where a fictitious brick wall (cutoff) is imposed at a short distance $h$ outside the horizon.
The brick wall model was proposed as an effective description of black holes,
such that, for example,
one can apply quantum mechanics 
to compute the entropy of a black hole
by imposing the boundary condition that
all wave functions of particles vanish on the wall.

A peculiar feature of the entropy formula (\ref{S}) is that
it is independent of the number of species of particles in the theory.
(In contrast, 
for instance,
the entropy of a box of gas depends on the number of different types of gases.)
As a result, 
in the brick wall model, 
in order to produce the correct black-hole entropy (\ref{S}),
the thickness of the brick wall has to be
proportional to the number of species of
massless particles:
\be
h = \frac{N_0 G}{720\pi M},
\ee
where $N_0$ is the number of species of light particles,
$G$ is the Newton constant
and $M$ is the mass of the black hole.
The peculiar feature $h \propto N_0$ 
is put in by hand in the brick wall model,
but it is automatically realized in the KMY model
because $R_0 - a_0 \simeq 2\sigma/R_0$ and $\sigma \propto N_0$.

In the KMY model,
the brick wall is replaced by the surface layer of the collapsing sphere
and it is real.
The wave functions of particles do not just vanish on the shell,
but can be extended into the sphere.
It is a concrete well-defined problem to compute 
the entropy of the system without imposing the cutoff by hand.
We leave this problem to the future.

\subsection{Membrane Paradigm}
\label{Membrane}

\begin{figure}
\vskip-2em
\center
\includegraphics[scale=0.4,bb=0 100 500 500]{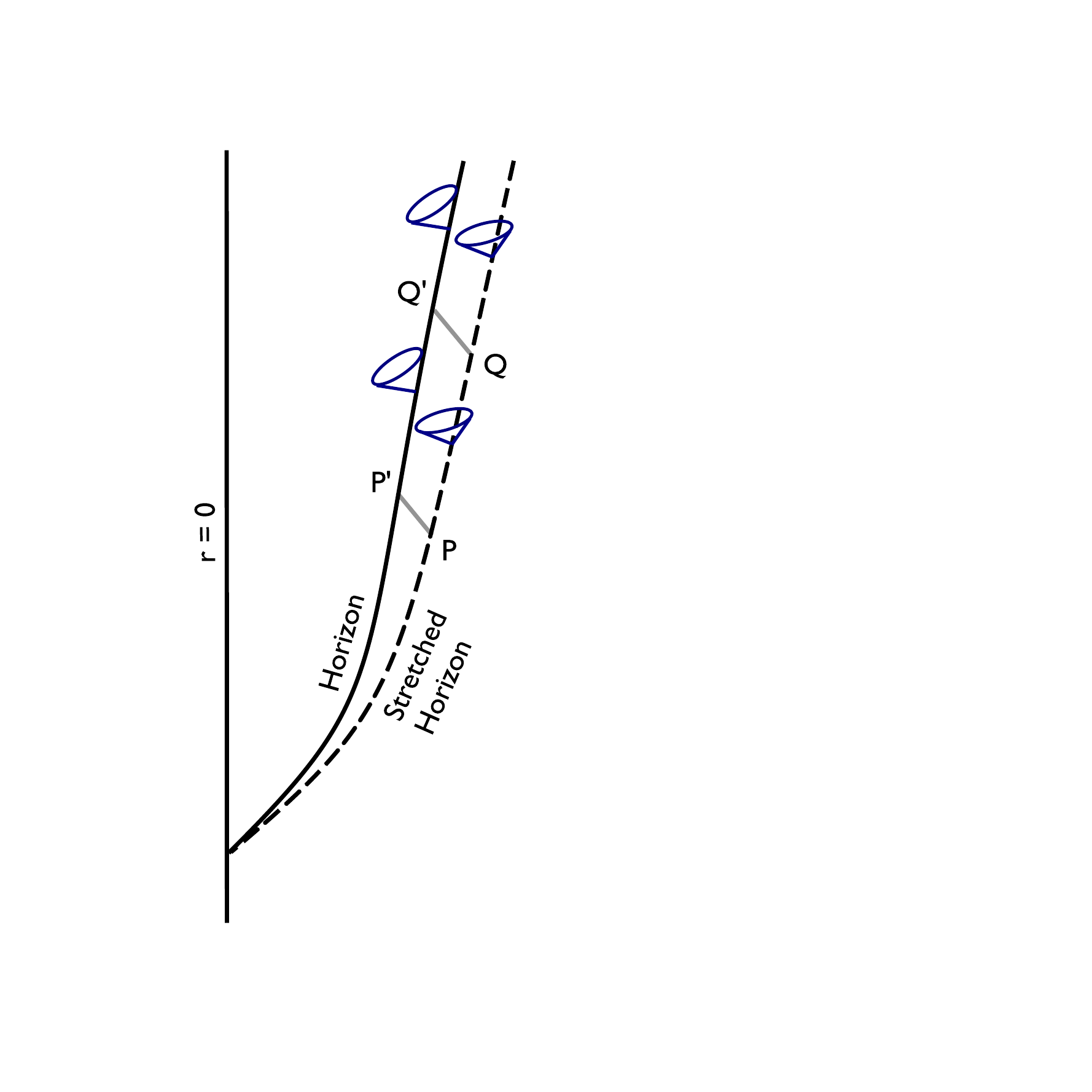}
\includegraphics[scale=0.4,bb=0 100 500 500]{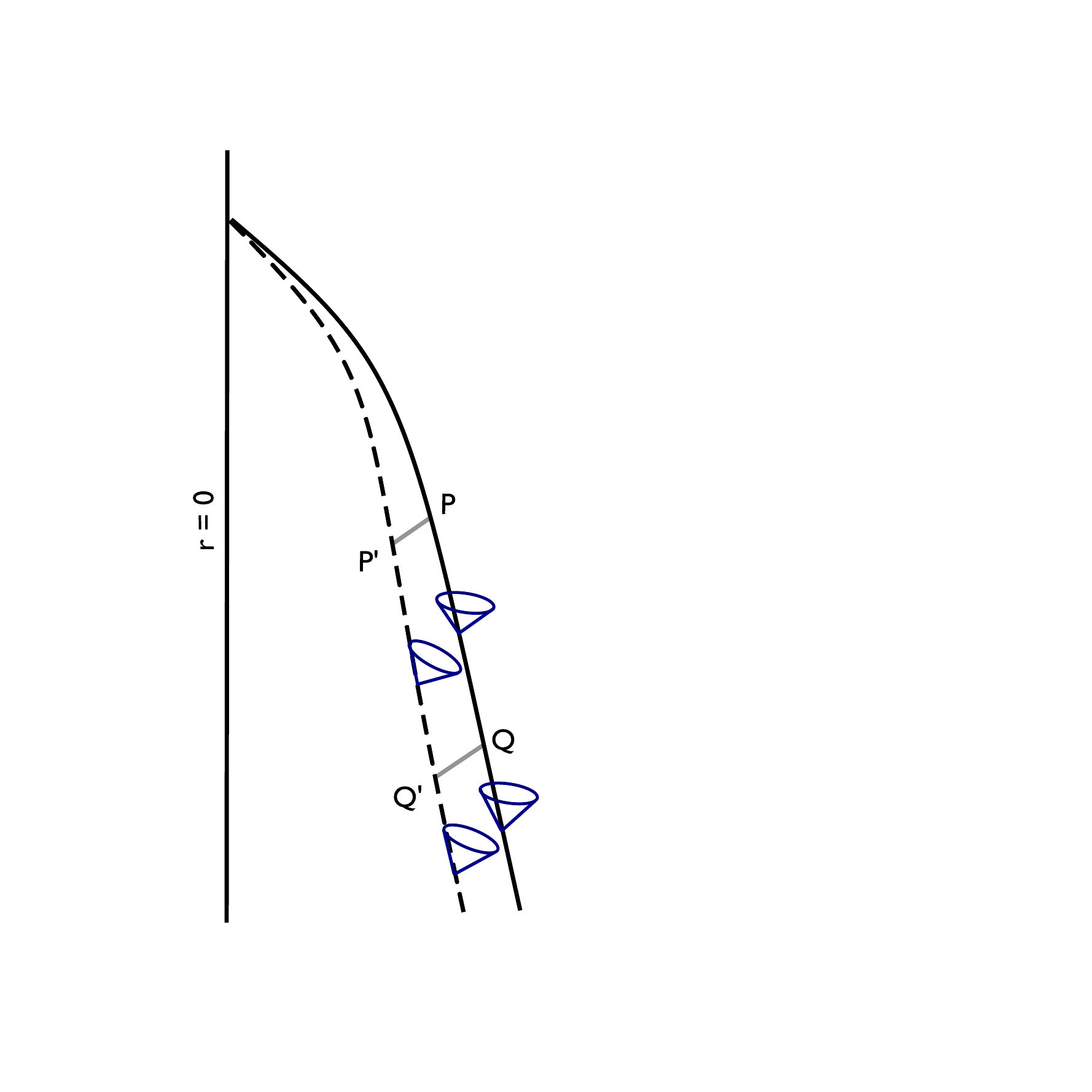}
\vskip2em
\hskip-4em
(a)\hskip15em(b)
\caption{\small
(a) In the membrane paradigm,
a stretched horizon (dash curve) lies outside 
the black-hole horizon (solid curve),
and every point (e.g. $P$, $Q$) on the stretched horizon 
has a corresponding point ($P'$, $Q'$) on the real horizon
connected by radial ingoing null lines.
(b) 
The surface of the collapsing body (solid line) 
lies outside the hypothetical white-hole horizon
(dash curve --- obtained by analytic continuation).
}
\label{Membrane-Paradigm}
\vskip1em
\end{figure}

The conventional membrane paradigm \cite{Thorne:1986iy} has 
a thin radiating surface outside the black-hole horizon
as an effective description of the black hole.
See Fig.\ref{Membrane-Paradigm}(a).
In the KMY model,
the surface layer of the collapsing matter acts like a physical membrane,
as we have discussed above.
If we analytically continue the outgoing Vaidya metric for $r > R_0$
into the space of $r < R_0$,
the surface of the collapsing matter is slightly outside the white-hole horizon.
See Fig.\ref{Membrane-Paradigm}(b).
Fig.\ref{Membrane-Paradigm}(b) differs from Fig.\ref{Membrane-Paradigm}(a)
not only by a time-reversal transformation 
but also a swap of the roles played by the membrane and the horizon:
which one is imaginary and which one is real.

What was said about the stretched horizon
in the membrane paradigm
needs some modification to apply to 
the surface layer of the asymptotic black hole.
For instance,
in the membrane paradigm,
the boundary condition of electromagnetic fields 
on the membrane is given by \cite{Price:1986yy}
\be
\vec{E} \times \hat{n} \simeq \vec{B},
\ee
where $\hat{n}$ is the (outward) unit normal vector of the membrane.
It says that,
in the limit when the membrane approaches to the horizon,
as the horizon is composed of outgoing light-like curves,
all electromagnetic fluctuations look like ingoing electromagnetic waves.

This condition should be reversed in the new picture.
The boundary condition of electromagnetic fields 
on the surface of the collapsing sphere should be
\footnote{
This assumes that there is nothing but outgoing radiation
outside the collapsing star.
It does not apply to an environment with CMB or other energy sources.
}
\be
\vec{E} \times \hat{n} \simeq - \vec{B},
\ee
saying that all electromagnetic fluctuations 
are outgoing waves.
It will be interesting to construct an effective theory 
for the collapsing sphere in which
physical degrees of freedom inside the sphere are ignored
except the thin layer (membrane) at the surface,
as the new membrane paradigm.

\subsection{Conclusion}

In this work,
we have derived the metric (\ref{metric}) for a collapsing matter sphere
including the back-reaction of an outgoing radiation 
as a solution to the Einstein equation.
The two parametric functions of the metric specify
the collapsing velocity $V(u, r)$ for the shell of radius $r$ at time $u$,
and the mass $m(u, r) = a(u, r)/2$ enclosed within this shell.
The red-shift factor $e^{\psi(u, r)}$ in the metric (\ref{metric})
is determined by these two functions through eq.(\ref{red-shift-general}).
The significant implication of this result is that 
it allows us to answer questions through explicit calculations,
rather than speculation or intuition.

The energy-momentum tensor for the metric (\ref{metric})
is decomposed into the contributions of three sources:
(1) infalling matter sphere,
(2) outgoing radiation,
and (3) pressure in the transverse directions.
We have checked that there is no $\delta$-function term
in the energy-momentum tensor at the surface of the collapsing sphere in general,
and the singularity at the origin can be avoided in the asymptotic black hole.

The special case of a sphere collapsing at the speed of light is considered in detail,
reproducing some of the results of Refs.\cite{Kawai:2013mda,Kawai:2014afa}.
The consideration of a generic gravitational collapse 
naturally leads to the notion of asymptotic black holes
as a universal asymptotic state for a wide range of initial conditions with spherical symmetry.
Similar notion was also mentioned in Refs.\cite{Kawai:2014afa,Kawai:2015uya}
without using the same terminology.

Despite our claim that a generic gravitational collapse
approaches to the state of an asymptotic black hole,
we do not have a rigorous statement about 
how general it is
or whether it is an attractor in the phase space.
In fact, 
as the notion of asymptotic black hole only applies to
large massive objects,
the very late stage of a gravitational collapse
shortly before the complete evaporation 
is expected to be different.

The exponential form of the red-shift factor (\ref{red-shift-aBH})
for an asymptotic black hole freezes up everything inside the collapsing sphere
except the outermost layer of thickness of order $\sqrt{2\sigma}$.
It effectively serves as a horizon
from the viewpoint of a distant observer
(see Secs.\ref{PointParticle}, \ref{r=0singularity}).
This feature is also reminiscent of both the brick wall model 
and the membrane paradigm.

\appendix

\section*{Appendix A: Ingoing and Outgoing Vaidya Metrics}
\label{metrics}

The Schwarzschild metric with a constant Schwarzschild radius $a_0$,
\be
ds^2 = - \left(1-\frac{a_0}{r}\right) dt^2 + \frac{1}{\left(1-\frac{a_0}{r}\right)} dr^2 + r^2 d\Omega^2,
\label{Schwarzschild-metric-static}
\ee
can also be expressed in terms of the Eddington advanced or retarded time $v$ or $u$ as
\bea
ds^2 &=& - \left(1-\frac{a_0}{r}\right) dv^2 + 2 dvdr + r^2 d\Omega^2,
\label{ingoing-Vaidya-metric-static}
\\
ds^2 &=& - \left(1-\frac{a_0}{r}\right) du^2 - 2 dudr + r^2 d\Omega^2,
\label{outgoing-Vaidya-metric-static}
\eea
where
\be
v \equiv t + r^* 
\qquad 
u \equiv t - r^*,
\ee
with
\be
r^* \equiv r + a_0\log\left(\frac{r}{a_0} - 1\right).
\ee

Turning on time dependence of the Schwarzschild radius,
the metrics above become
\bea
ds^2 &=& - \left(1-\frac{a_S(t)}{r}\right) dt^2 + \frac{1}{\left(1-\frac{a_S(t)}{r}\right)} dr^2 + r^2 d\Omega^2,
\label{Schwarzschild-metric}
\\
ds^2 &=& - \left(1-\frac{a_I(v)}{r}\right) dv^2 + 2 dvdr + r^2 d\Omega^2,
\label{ingoing-Vaidya-metric}
\\
ds^2 &=& - \left(1-\frac{a_O(u)}{r}\right) du^2 - 2 dudr + r^2 d\Omega^2,
\label{outgoing-Vaidya-metric}
\eea
The energy-momentum tensors of the ingoing and outgoing Vaidya metrics
(\ref{ingoing-Vaidya-metric}), (\ref{outgoing-Vaidya-metric}) are
\bea
T_{vv} &=& \frac{1}{8\pi G} \frac{\dot{a}_I(v)}{r^2},
\label{EMT-ingoing-Vaidya}
\\
T_{uu} &=& - \frac{1}{8\pi G} \frac{\dot{a}_O(u)}{r^2},
\label{EMT-outgoing-Vaidya}
\eea
respectively,
with $T_{vr} = T_{rr} = 0$.

The ingoing Vaidya metric (\ref{ingoing-Vaidya-metric})
satisfies the weak energy condition only if 
\be
\dot{a}_I(v) \geq 0.
\label{increasing-a}
\ee
It can describe the interior of a sphere of infalling massless dust,
with the infalling energy-momentum tensor $T_{vv}$ (\ref{EMT-ingoing-Vaidya})
(without Hawking radiation).
The outgoing Vaidya metric (\ref{outgoing-Vaidya-metric})
satisfies the weak energy condition only if
\be
\dot{a}_O(u) \leq 0.
\label{decreasing-a}
\ee
It can describe outgoing radiation
with the energy-momentum tensor $T_{uu}$ (\ref{EMT-outgoing-Vaidya})
outside a radiating star.

Both of the ingoing and outgoing Vaidya metrics (\ref{ingoing-Vaidya-metric}), (\ref{outgoing-Vaidya-metric})
can be viewed as small perturbations of the Schwarzschild metric
if $a_I(v)$ and $a_O(u)$ appear to be approximately constant.
For a sufficiently small time-derivative ($|\dot{a}_I(v)|$ or $|\dot{a}_O(u)|$)
the space-time region outside the Schwarzschild radius $a_I(v)$ or $a_O(u)$
can be well approximated by the Schwarzschild metric
over an extremely long period of time for a distant observer.
It can be very difficult to distinguish the three metrics:
the Schwarzschild metric (no radiation),
the ingoing Vaidya metric (ingoing radiation) 
and the outgoing Vaidya metric (outgoing radiation)
if the radiation is too weak to be detected.
In all three cases,
there is a large red-shift factor so that all infalling matter 
would appear to be motion-less as it approaches to the horizon
from the viewpoint of a distant observer.

Nevertheless,
however small the derivative $\dot{a}_O$ is,
as long as it is finite and negative
($\dot{a}_0 < - \eps$ for fixed $\eps > 0$),
eventually $a_O$ goes to zero,
so that $a_O(u) = 0$ for all $u \geq u^*$ for some finite $u^*$.
At any finite time $u < u^*$,
the radiation can be so weak that a distant observer 
cannot tell its difference from a Schwarzschild black hole,
yet all the energy evaporates completely within finite time
so that the outgoing Vaidya metric turns into the Minkowski space.
All time-like and light-like ingoing particles originated outside $a_O$
can never cross inside the Schwarzschild radius $a_O$.
This feature is completely different from 
the case of an ingoing Vaidya metric,
for which ingoing particles can pass through the Schwarzschild radius $a_I$
within finite proper time.

In fact,
while the surface at $r = a_I(v)$ is a trapping surface 
(constant-$r$ curves are outgoing light-like curves)
for the ingoing Vaidya metric,
the surface at $r = a_O(u)$ is an ``untrapping surface''
(constant-$r$ curves are ingoing light-like curves)
for the outgoing Vaidya metric.
We refer to the trapping surface as the black-hole (apparent) horizon,
and the ``untrapping surface'' as the white-hole (apparent) horizon.

The sphere at $r = a_0$ in the Schwarzschild metric
can be either a black-hole horizon or a white-hole horizon.
The two horizons are degenerate in the coordinate $r$.
An important implication of the degeneracy of the black hole and white hole
in the Schwarzschild solution is that
a small deformation of the Schwarzschild solution 
can be dramatically different from another small deformation.

Another way to look at this feature is to say that
a small deformation in one reference frame may not be small 
in another reference frame.
Indeed,
gravitational collapses are critical phenomena
\cite{CriticalPhenomena}.
An infinitesimal change in the initial conditions
may lead to a change from the black-hole state with a horizon
to a horizon-less configuration.

\section*{Appendix B: Hawking Radiation}

In general, 
a nontrivial Bogoliubov transformation
due to a time-dependent background 
can lead to the creation of particles.
It has been shown that 
a Hawking-like radiation at temperature $T_H$ is created
whenever the affine parameters $U$ and $u$
for the past and future infinities
(which are assumed to be asymptotically flat)
satisfy the relation \cite{HR2}
\be
U \simeq U_0 - A_0 e^{-\kappa_H u}
\label{U-u-1}
\ee
approximately for some constants $U_0$, $A_0$ and $\kappa_H$.
The constant $\kappa_H$ determines the temperature of Hawking-like radiation:
\be
T_H = \frac{\kappa_H}{2\pi}.
\label{TH}
\ee
More precisely,
we need to check that 
the exponential relation (\ref{U-u-1}) holds approximately 
over a sufficiently large range of space-time
in order for a good approximation of low frequency modes.
In general, 
ingoing light-like curves reflects at the origin to turn into outgoing light curves,
so that $U$ can be viewed as a function of $u$.
One can always define
\be
\kappa(u) \equiv - \frac{\frac{d^2 U}{du^2}}{\frac{dU}{du}}
= - \frac{d}{du} \log\left(\frac{dU}{du}\right),
\label{kappa-1}
\ee
and the question is whether $\kappa(u)$ 
is approximately a constant.
Define
\be
{\cal D} \equiv \mbox{sup}_{n > 0} 
\left\{ \frac{1}{(n+1)!} \frac{|\kappa^{(n)}|}{\kappa^{n+1}} \right\},
\label{D}
\ee
where $\kappa^{(n)}$ is the $n$-th derivative of $\kappa(u)$
with respect to $u$.
Then the exponential relation (\ref{U-u-1}) is a good approximation for
\be
|u - u_H| \ll \frac{1}{\sqrt{2}{\cal D}\kappa_H}
\label{u-u_H}
\ee
where $u_H$ is a moment when $\kappa(u_H) = \kappa_H$.

For the KMY model,
one has to slightly modify the calculation above so that
eq. (\ref{U-u-1}) is replaced by
\be
U \simeq U_0 + A_0 e^{\kappa_H u},
\label{U-u-2}
\ee
where there are two changes of sign.
A change in $u$ leads to an exponentially small or large change in $U$
according to eq.(\ref{U-u-1}) or (\ref{U-u-2}).
This is not unexpected as
the white hole is a time-reversal transform of the black hole.
Accordingly,
the Bogoliubov transformation coefficients for the exponential relation (\ref{U-u-2})
are the complex conjugation of those for eq.(\ref{U-u-1}) \cite{Kawai:2013mda}.
However, 
the probability of particle creation is the absolute value squared 
of the Bogoliubov transformation coefficients.
As a result,
the same spectrum of radiation is predicted for 
both exponential relations (\ref{U-u-1}) and (\ref{U-u-2}).

The definition of $\kappa$ (\ref{kappa-1})
should now be modified to
\be
\kappa(u) \equiv \frac{\frac{d^2 U}{du^2}}{\frac{dU}{du}}
= \frac{d}{du} \log\left(\frac{dU}{du}\right),
\label{kappa-2}
\ee
so that (\ref{D}) and (\ref{u-u_H}) remain unchanged.
For an asymptotic black hole,
we have \cite{Kawai:2013mda}
\be
\kappa(u) \simeq \frac{1}{2a_0}
\ee
for large $a_0$,
according to eq.(\ref{huge-red-shift}).
Combined with eq.(\ref{a-trajectory})
we find that ${\cal D}$ (\ref{D}) is dominated by the case of $n=1$,
so that
$2{\cal D}^2 \simeq \frac{|\dot{\kappa}|}{\kappa^2}$.
Thus, 
eq.(\ref{u-u_H}) implies that
the exponential approximation (\ref{U-u-2}) holds for
\be
|u-u_H| \ll |\dot{\kappa}|^{-1/2} \simeq \sqrt{\frac{2}{\sigma}} \, a_0^2.
\ee
The thermal radiation with the temperature $T_H$ (\ref{TH})
is thus a good approximation of the radiation of the KMY model
for wave-lengths much shorter than $a_0^2/\sqrt{\sigma}$.
As the pre-Hawking radiation is dominated by wave-lengths of order $a_0$
(for a distant observer),
the thermal spectrum is a good approximation for $a_0 \gg \sqrt{\sigma}$.

The total power of radiation can also be calculated independently
as the vacuum expectation value of the energy-momentum tensor
via the point-slitting regularization.
The result is \cite{Kawai:2013mda}
\be
\langle T_{uu} \rangle = \frac{1}{4\pi r^2} \frac{N}{8\pi}\{u, U\},
\ee
where the bracket
is the Schwarzian derivative (\ref{Schwarzian}).
This implies that the Schwarzschild radius $a_0$ decreases with time as 
\cite{Kawai:2013mda}
\be
\dot{a}_0(u) = \frac{N \ell_p^2}{4\pi}\{u, U\}.
\label{dot-a}
\ee
This is the origin of eq.(\ref{dot-a-a}).

\section*{Appendix C: Infalling Observer's Perspective}

Although it is mathematically proven that 
there would be no horizon 
whenever there is complete evaporation,
regardless of how long it takes to evaporate,
one may find it perplexing why an infalling observer,
as he falls in with the collapsing sphere,
could see a massive star totally evaporate
into quantum particles 
before the horizon emerges.

Let us calculate the proper time for the infalling observer 
to see the collapsing star evaporate away.
Assuming that the infalling observer can be described 
as a particle of mass $m$ in free fall,
its trajectory can be described by the action
\be
S \equiv - m \int d\tau
= -m \int du \; \sqrt{1-\frac{a_0(u)}{r(u)}+2\dot{r}(u)},
\ee
where $\tau$ is the proper time for the infalling observer,
and its equation of motion is
\be
\frac{d}{du}\log(K(u)) = - \frac{1}{2} \frac{a_0(u)}{r^2(u)},
\label{eom-K}
\ee
where
\be
K(u) \equiv \frac{d\tau}{du}
= \sqrt{1 - \frac{a_0(u)}{r(u)} + 2\dot{r}(u)}.
\ee

Apparently,
$K^2(u)$ has to be non-negative in order for the Lagrangian to be real.
(It vanishes when the observer moves at the speed of light.)
For an infalling process, 
$\dot{r}(u) < 0$,
hence $r(u) > a_0(u)$.
If $r(u)$ approaches to $a_0(u)$ as $u \rightarrow \infty$
but within finite proper time ($\int_0^{\infty} d\tau < \infty$),
the trajectory of $r(u)$ is geodesically incomplete,
and it means that there is a horizon at $u = \infty$.
The other possibility is that
$a_0(u)$ goes to $0$ at finite $u$
(e.g. $a_0(u) = 0$ for $u \geq u^*$).
Then $r(u)$ goes to $0$ at some $\bar{u} > u^*$,
and it is geodesically complete 
as it continues in Minkowski space for $u \geq \bar{u}$.
We shall consider the latter case when
$a_0(u)$ is a solution to eq.(\ref{a-trajectory})
given by eq.(\ref{a-u-1/3}).
Denote the mass of the star at $u = 0$ by $M$,
then $u^* = (2GM)^3/(3\sigma)$.

For $r(u)$ close to $a(u)$, 
the right hand side of eq.(\ref{eom-K}) is roughly $-\frac{1}{2a_0(u)}$
(this approximation would underestimate the proper time),
and so
\be
K(u) \simeq K(0) e^{- \int_0^u \frac{du'}{2a_0(u')}}
\ee
for a constant $K(0)$,
which is related to the initial condition of $r(0)$ and $\dot{r}(0)$ by
\be
K(0) \simeq \sqrt{1- \frac{a_0(0)}{r(0)} + 2\dot{r}(0)}.
\ee
An estimate of $K(u)$ is thus
\be
K(u) \simeq K(0) e^{-\frac{3}{4(3\sigma)^{1/3}}(u^*{}^{2/3} - (u^*-u)^{2/3})}.
\ee

In terms of $K(u)$ as a measure of how close the observer is to the speed of light,
this expression tells us that 
the observer approaches to the speed of light exponentially 
over a time scale of order $(u^*\sigma)^{1/3}$.
The proper time for the infalling observer to see 
the complete evaporation of the star of initial mass $M$ is approximately
\be
T \equiv \int d\tau = \int_0^{u^*} du K(u) \simeq 4GM K(0) + {\cal O}(1/M),
\ee
where we have ignored subleading terms in the large-$M$ expansion.
If we take the limit of constant $a_0$ 
(correspondingly, $u^*\rightarrow\infty$),
the value of $T$ remains roughly the same.
(The proper time of free fall is not dramatically changed
by the pre-Hawking radiation.)
The difference is that,
if $a_0$ is a constant,
the infalling observer falls to the horizon at $r = a_0$ 
within finite proper time $T$ 
(as $u \rightarrow \infty$).
The infalling trajectory is geodesically incomplete,
and we know that there is a horizon at $u = \infty$.

Here we consider the case of $a_0 \rightarrow 0$ at finite $u^*$ (\ref{CE}).
The question is whether the proper time $T \sim 4GM$
($K(0)$ is always less than $1$)
is long enough to see the whole star evaporate away.
Hence we shall calculate the energy flux in pre-Hawking radiation 
from the viewpoint of the infalling observer.
The energy-momentum tensor (\ref{EMT-outgoing-Vaidya-0}) 
for pre-Hawking radiation is interpreted by the infalling observer
as a radiation of power
\be
{\cal P} = \int d^2\Omega \, T_{uu} \left(\frac{du}{d\tau}\right)^2
\simeq 
\frac{1}{2G} \frac{\sigma}{a_0^2} \left(\frac{du}{d\tau}\right)^2,
\label{calP}
\ee
assuming that the infalling observer is close to 
the surface $R_0$ of the collapsing sphere,
which is close to the Schwarzschild radius $a_0$.
(We have also used eq.(\ref{a-trajectory}).)

The energy flux ${\cal P}$ appears to be very small due to the factor of $1/a_0^2$,
despite the large blue-shift factor $1/(1-a_0/r)$ involved in the calculation
\cite{Ho:2015vga}.
It is large only if the observer is falling close to the speed of light
($du/d\tau$ diverges at the speed of light).
In fact,
the energy flux that should be integrated over time 
to account for the evaporation of the Bondi mass $M$ is even smaller,
given by
\be
{\cal P}' \equiv \frac{d\tau}{du} {\cal P} 
= \frac{1}{2G} \frac{\sigma}{a_0^2}\frac{du}{d\tau},
\label{Pprime}
\ee
because the Bondi mass is defined with respect to the time coordinate $u$.

Naively, 
there is a mismatch in the total time $T$ of evaporation (of order $M$)
and the energy flux ${\cal P}'$ (\ref{Pprime})
(naively of order $1/M^2$),
as their product should be of order $M$
for complete evaporation.
Nevertheless,
it can be checked that
the integral of radiation power over proper time
\be
\int_0^T d\tau \, {\cal P}' \simeq M
\ee
indeed gives the correct Bondi mass.
This is because the infalling observer
is accelerated by gravitational force towards the speed of light,
and the factor $du/d\tau$ in ${\cal P}'$ can be very large.
This is the same effect as a Lorentz boost.
If you move in the opposite direction of any radiation
at the speed of light,
the energy flux in the radiation would appear to be infinite.
(The power ${\cal P}'$ of radiation is very small before
the infalling observer is accelerated to produce a large factor of $1/K(u)$,
regardless of how close he is to the collapsing star.)

\section*{Appendix D: Metric of KMY Model}

A phenomenological expression for the metric of a collapsing body 
is given in Ref.\cite{Kawai:2014afa} 
via three parametric functions $a_0(u)$, $f(r)$ and $\sigma(r)$ as
\footnote{
The notation here is slightly different from Ref.\cite{Kawai:2014afa},
where $B$ is denoted as $B^{-1}$.
}
\be
ds^2 = - e^{-A(u, r)/2}
\left(B(r) e^{-A(u, r)/2} du + 2 dr\right) du + r^2 d\Omega^2,
\ee
where
\bea
B(r) &\simeq& \left\{
\begin{array}{ll}
\frac{2\sigma(r)}{r^2}
& (r \leq R_0(u)), \\
1-\frac{a_0(u)}{r}
& (r \geq R_0(u)),
\end{array}
\right.
\\
A(u, r) &\simeq& \left\{
\begin{array}{ll}
\int_r^{R_0(u)} dr' \frac{r'}{(1+f(r'))\sigma(r')}
& (r \leq R_0(u)), \\
0 
& (r \geq R_0(u)),
\end{array}
\right.
\label{A-def}
\eea
and the radial coordinate of the surface of the collapsing sphere is approximately
\be
R_0(u) \simeq a_0(u) + \frac{2\sigma(R_0(u), u)}{a_0(u)},
\ee
These expressions are good approximations for large $a_0^2/\sigma$.

The function $a_0(u)$ is the Schwarzschild radius
for the outgoing Vaidya metric outside the collapsing sphere.
The function $f(r)$ specifies the strength of scattering of the outgoing radiation
from the interior of the collapsing body.
If there is no scattering, 
$f(r) = 0$.
The function $\sigma(r)$ specifies how fast 
the mass of the collapsing body reduces over time 
due to pre-Hawking radiation.
For a large asymptotic black hole,
it is approximately a constant (\ref{sigma})
for large $r$.

For a quasi-black hole in equilibrium with the environment,
the metric is \cite{Kawai:2014afa}:
\be
ds^2 = - e^{-A(r)} B(r) dt^2 + \frac{1}{B(r)} dr^2 + r^2 d\Omega^2,
\ee
where $A(r)$ is the same function $A(u, r)$ (\ref{A-def})
but with $R_0$ being a time-independent constant.

\section*{Acknowledgement}

The author would like to thank 
Hikaru Kawai for sharing his original ideas
on which this work is based,
and to thank
Dongsu Bak,
Heng-Yu Chen, Yi-Chun Chin, Chong-Sun Chu, Bartek Czech, 
Gary Horowitz, Kazuo Hosomichi, Yu-tin Huang, 
Takeo Inami, Hsien-chung Kao, Shinsuke Kawai, 
Dieter L\"{u}st, Yutaka Matsuo, 
Yen Chin Ong, Gary Shiu,
Mu-Tao Wang, Wen-Yu Wen, 
Shu-Jung Yang, Dong-han Yeom, Piljin Yi, Xi Yin and Zheng Yin
for discussions.
The work is supported in part by
the Ministry of Science and Technology, R.O.C.
(project no. 104-2112-M-002 -003 -MY3)
and by National Taiwan University
(project no. 105R8700-2).


\end{CJK} 

\vskip .8cm
\baselineskip 22pt
\begin{thebibliography}{99}
\itemsep 0pt

\bibitem{Kawai:2013mda} 
  H.~Kawai, Y.~Matsuo and Y.~Yokokura,
  ``A Self-consistent Model of the Black Hole Evaporation,''
  Int.\ J.\ Mod.\ Phys.\ A {\bf 28}, 1350050 (2013)
  [arXiv:1302.4733 [hep-th]].
  
\bibitem{Kawai:2014afa} 
  H.~Kawai and Y.~Yokokura,
  ``Phenomenological Description of the Interior of the Schwarzschild Black Hole,''
  arXiv:1409.5784 [hep-th].

\bibitem{Ho:2015fja} 
  P.~M.~Ho,
  ``Comment on Self-Consistent Model of Black Hole Formation and Evaporation,''
  arXiv:1505.02468 [hep-th].

\bibitem{Kawai:2015uya} 
  H.~Kawai and Y.~Yokokura,
  ``Interior of Black Holes and Information Recovery,''
  arXiv:1509.08472 [hep-th].

\bibitem{Ho:2015vga} 
  P.~M.~Ho,
  ``The Absence of Horizon in Black-Hole Formation,''
  arXiv:1510.07157 [hep-th].

\bibitem{Vachaspati:2006ki} 
  T.~Vachaspati, D.~Stojkovic and L.~M.~Krauss,
  ``Observation of incipient black holes and the information loss problem,''
  Phys.\ Rev.\ D {\bf 76}, 024005 (2007)
  [gr-qc/0609024].

\bibitem{Saini:2015dea} 
  A.~Saini and D.~Stojkovic,
  ``Radiation from a collapsing object is manifestly unitary,''
  Phys.\ Rev.\ Lett.\  {\bf 114}, no. 11, 111301 (2015)
  [arXiv:1503.01487 [gr-qc]].

\bibitem{Mersini-Houghton:2014zka} 
  L.~Mersini-Houghton,
  ``Backreaction of Hawking Radiation on a Gravitationally Collapsing Star I: Black Holes?,''
  PLB30496 Phys Lett B, 16 September 2014
  [arXiv:1406.1525 [hep-th]].

\bibitem{Mersini-Houghton:2014cta} 
  L.~Mersini-Houghton and H.~P.~Pfeiffer,
  ``Back-reaction of the Hawking radiation flux on a gravitationally collapsing star II: Fireworks instead of firewalls,''
  arXiv:1409.1837 [hep-th].

\bibitem{Almheiri:2012rt} 
  A.~Almheiri, D.~Marolf, J.~Polchinski and J.~Sully,
  ``Black Holes: Complementarity or Firewalls?,''
  JHEP {\bf 1302}, 062 (2013)
  [arXiv:1207.3123 [hep-th]];
  
\bibitem{Braunstein}
S. L. Braunstein, 
``Black hole entropy as entropy of entanglement, 
  or it's curtains for the equivalence principle,''
[arXiv:0907.1190v1 [quant-ph]] 
published as 
  S.~L.~Braunstein, S.~Pirandola and K.~Życzkowski,
  ``Better Late than Never: Information Retrieval from Black Holes,''
  Phys.\ Rev.\ Lett.\  {\bf 110}, no. 10, 101301 (2013),
  for a similar prediction from different assumptions.

\bibitem{Vaidya:1951zz} 
  P.~Vaidya,
  ``The Gravitational Field of a Radiating Star,''
  Proc.\ Indian Acad.\ Sci.\ A {\bf 33}, 264 (1951).

\bibitem{mass}
   T. Zannias, Phys.\ Rev.\ D {\bf 41}, 3252 (1990);
   E. Poisson, W. Israel, Phys.\ Rev.\ D{\bf 41}, 1976 (1990).

\bibitem{Hawking:1974sw} 
  S.~W.~Hawking,
  ``Particle Creation by Black Holes,''
  Commun.\ Math.\ Phys.\  {\bf 43}, 199 (1975)
  [Commun.\ Math.\ Phys.\  {\bf 46}, 206 (1976)].
  S.~W.~Hawking,
  ``Breakdown of Predictability in Gravitational Collapse,''
  Phys.\ Rev.\ D {\bf 14}, 2460 (1976).

\bibitem{HR1}
  P.~Hajicek,
  ``On the Origin of Hawking Radiation,''
  Phys.\ Rev.\ D {\bf 36}, 1065 (1987).
  M.~Visser,
  ``Essential and inessential features of Hawking radiation,''
  Int.\ J.\ Mod.\ Phys.\ D {\bf 12}, 649 (2003)
  [hep-th/0106111].
  C.~Barcelo, S.~Liberati, S.~Sonego and M.~Visser,
  ``Quasi-particle creation by analogue black holes,''
  Class.\ Quant.\ Grav.\  {\bf 23}, 5341 (2006)
  [gr-qc/0604058].
  C.~Barcelo, S.~Liberati, S.~Sonego and M.~Visser,
  ``Hawking-like radiation does not require a trapped region,''
  Phys.\ Rev.\ Lett.\  {\bf 97}, 171301 (2006)
  [gr-qc/0607008].
  
\bibitem{HR2}
  C.~Barcelo, S.~Liberati, S.~Sonego and M.~Visser,
  ``Minimal conditions for the existence of a Hawking-like flux,''
  Phys.\ Rev.\ D {\bf 83}, 041501 (2011)
  [arXiv:1011.5593 [gr-qc]].
  C.~Barcelo, S.~Liberati, S.~Sonego and M.~Visser,
  ``Hawking-like radiation from evolving black holes and compact horizonless objects,''
  JHEP {\bf 1102}, 003 (2011)
  [arXiv:1011.5911 [gr-qc]].

\bibitem{Unruh:1977ga} 
  W.~G.~Unruh,
  ``Origin of the Particles in Black Hole Evaporation,''
  Phys.\ Rev.\ D {\bf 15}, 365 (1977).
  doi:10.1103/PhysRevD.15.365

\bibitem{Mathur:2009hf} 
  S.~D.~Mathur,
  ``The Information paradox: A Pedagogical introduction,''
  Class.\ Quant.\ Grav.\  {\bf 26}, 224001 (2009)
  [arXiv:0909.1038 [hep-th]].

\bibitem{Banks:2010zn} 
  T.~Banks and N.~Seiberg,
  ``Symmetries and Strings in Field Theory and Gravity,''
  Phys.\ Rev.\ D {\bf 83}, 084019 (2011)
  doi:10.1103/PhysRevD.83.084019
  [arXiv:1011.5120 [hep-th]].

\bibitem{FuzzBall}
  O.~Lunin and S.~D.~Mathur,
  ``AdS / CFT duality and the black hole information paradox,''
  Nucl.\ Phys.\ B {\bf 623}, 342 (2002)
  [hep-th/0109154].
  O.~Lunin and S.~D.~Mathur,
  ``Statistical interpretation of Bekenstein entropy for systems with a stretched horizon,''
  Phys.\ Rev.\ Lett.\  {\bf 88}, 211303 (2002)
  [hep-th/0202072].

\bibitem{'tHooft:1984re} 
  G.~'t Hooft,
  ``On the Quantum Structure of a Black Hole,''
  Nucl.\ Phys.\ B {\bf 256}, 727 (1985).

\bibitem{'tHooft:1996tq} 
  G.~'t Hooft,
  ``The Scattering matrix approach for the quantum black hole: An Overview,''
  Int.\ J.\ Mod.\ Phys.\ A {\bf 11}, 4623 (1996)
  doi:10.1142/S0217751X96002145
  [gr-qc/9607022].

\bibitem{Thorne:1986iy} 
  K.~S.~Thorne, R.~H.~Price and D.~A.~Macdonald,
  ``Black Holes: The Membrane Paradigm,''
  NEW HAVEN, USA: YALE UNIV. PR. (1986) 367p

\bibitem{Price:1986yy} 
  R.~H.~Price and K.~S.~Thorne,
  ``Membrane Viewpoint on Black Holes: Properties and Evolution of the Stretched Horizon,''
  Phys.\ Rev.\ D {\bf 33}, 915 (1986).
  doi:10.1103/PhysRevD.33.915

\bibitem{CriticalPhenomena}
  M.~W.~Choptuik,
  ``Universality and scaling in gravitational collapse of a massless scalar field,''
  Phys.\ Rev.\ Lett.\  {\bf 70}, 9 (1993).
  doi:10.1103/PhysRevLett.70.9
  C.~Gundlach,
  ``Critical phenomena in gravitational collapse,''
  Adv.\ Theor.\ Math.\ Phys.\  {\bf 2}, 1 (1998)
  [gr-qc/9712084].
For a review, see:
  C.~Gundlach and J.~M.~Martin-Garcia,
  ``Critical phenomena in gravitational collapse,''
  Living Rev.\ Rel.\  {\bf 10}, 5 (2007)
  doi:10.12942/lrr-2007-5
  [arXiv:0711.4620 [gr-qc]].







\end{thebibliography}
\end{document}